\documentclass[fleqn,usenatbib]{mnras}

\usepackage{ae,aecompl}
\usepackage{graphicx}
\usepackage[dvipsnames]{xcolor}
\usepackage{amsmath}
\usepackage{amssymb}
\usepackage{savesym}
\savesymbol{tablenum}
\usepackage{makecell}
\usepackage{siunitx}
\usepackage{xspace}
\usepackage{listings}
\usepackage{soul}
\usepackage{float}
\restoresymbol{SIX}{tablenum}

\usepackage{comment}

\usepackage[T1]{fontenc}
\usepackage{fontawesome}
\definecolor{linkcolor}{HTML}{2AAD2E}

\usepackage{newtxtext,newtxmath}

\newcommand{\fr}{\texttt{frank}\xspace}
\newcommand{\gal}{\texttt{galario}\xspace}
\newcommand{\cl}{\texttt{CLEAN}\xspace} 
\newcommand{\ws}{$w_{\rm smooth}$\xspace}
\newcommand{\al}{$\alpha$\xspace}
\newcommand{\ml}{M$\lambda$\xspace}
\newcommand{\kl}{k$\lambda$\xspace}

\title[Super-resolution trends in the ALMA Taurus survey]{Super-resolution trends in the ALMA Taurus survey: Structured inner discs and compact discs}

\author[Jennings, Tazzari, Clarke, Booth, Rosotti]{
Jeff Jennings,$^1$\thanks{E-mail: jmj51@ast.cam.ac.uk} Marco Tazzari,$^1$ Cathie J. Clarke,$^1$ Richard A. Booth,$^2$ and \newauthor Giovanni P. Rosotti$^{3, 4}$
\\
$^1$Institute of Astronomy, University of Cambridge, Madingley Road, Cambridge, CB3 0HA, UK \\
$^2$Astrophysics Group, Imperial College London, Prince Consort Road, London SW7 2AZ, UK \\
$^3$Leiden University, Niels Bohrweg 2, NL-2333 CA Leiden, Netherlands \\
$^4$School of Physics and Astronomy, University of Leicester, Leicester LE1 7RH, UK
}

\date{Accepted XXX. Received YYY; in original form ZZZ}

\pubyear{2022}

\begin{document}
\label{firstpage}
\pagerange{\pageref{firstpage}--\pageref{lastpage}}
\maketitle

\begin{abstract}
The $1.33$~mm survey of protoplanetary discs in the Taurus molecular cloud found annular gaps and rings to be common in extended sources ($\gtrsim 55$~au), when their 1D visibility distributions were fit parametrically. 
We first demonstrate the advantages and limitations of {\it nonparametric} visibility fits for data at the survey's $0.12 \arcsec$ resolution. Then we use the nonparametric model in Frankenstein
(\fr) to identify new substructure in three compact and seven extended sources. Among the new features we identify three trends: 
a higher occurrence rate of substructure in the survey's compact discs than previously seen,
underresolved (potentially azimuthally asymmetric) substructure in the innermost disc of extended sources,
and a \lq{}shoulder\rq{} on the trailing edge of a ring in discs with strong depletion at small radii.
Noting the shoulder morphology is present in multiple discs observed at higher resolution, we postulate it is tracing a common physical mechanism. We further demonstrate how a super-resolution \fr brightness profile is useful in motivating an accurate parametric model, using the highly structured source DL~Tau in which \fr finds two new rings. Finally we show that sparse $(u, v)$ plane sampling may be masking the presence of substructure in several additional compact survey sources.
\end{abstract}


\section{Background}
\label{sec:intro}
Numerous physical mechanisms are capable of producing axisymmetric (and in many cases also asymmetric) substructures in protoplanetary discs, with the list of candidates growing. Categories include forming and newly formed planets \citep[e.g.,][]{1979ApJ...233..857G, 1986ApJ...309..846L, 2012ARA&A..50..211K}; opacity effects due to ice sublimation fronts \citep[e.g.,][]{2016ApJ...818L..16Z, 2016ApJ...821...82O, 2019ApJ...885...36H}; gas-dust coupling effects, including preferential dust growth in localized regions \citep[e.g.,][]{2012A&A...538A.114P, 2018ApJ...869L..46D, 2019ApJ...876....7S}, gravitational instability \citep[e.g.,][]{2015MNRAS.451..974D, 2018MNRAS.477.1004H, Dong_2018}, dynamical effects of a central binary \citep[e.g.,][]{2017MNRAS.464.1449R, 10.1093/mnras/sty647, 2021MNRAS.503.4930L}, and internal photoevaporation \citep[e.g.,][]{Clarke2001, Alexander_2006, Ercolano_2009}; and magnetic field effects including dead-zone boundaries \citep[e.g.,][]{2006A&A...446L..13V, Flock2015, 2016A&A...596A..81P}, magnetic flux concentration and zonal flows \citep[e.g.,][]{Johansen_2009, Bai_2014, 2021arXiv210610167C}, and the vertical shear instability \citep[e.g.,][]{2017ApJ...850..131F, 2018MNRAS.480.2125M, 2020arXiv200811195P}.

Determining which of these mechanisms dominate in observed systems requires both in-depth studies of individual sources and a large ensemble of discs with characterized substructure. Interferometric observations offer the highest spatial resolution to characterize disc features, and numerous works at the best resolutions achieved to-date in the (sub-)mm, $\approx 25 - 75$~mas ($\approx 1 - 10$~au), have confirmed (along with many critical works at moderate resolution) that the $\approx$~mm dust distribution in these discs is commonly structured. At these high resolutions,
the DSHARP \citep{Andrews2018, 2018ApJ...869L..42H} and ODISEA \citep{2021MNRAS.501.2934C} surveys, as well as several high resolution case studies of individual systems \citep[e.g.,][]{Partnership2015a, Andrews2016, 2018ApJ...857...18S, 2018ApJ...860..124D, 2018ApJ...866L...6C, 2018ApJ...868L...5K, Keppler_2019, 2019NatAs.tmp..419P, 2019arXiv190205143P, 2019ApJ...878L...8T, 2020ApJ...891...48H, 2021A&A...648A..33M, 2021ApJ...911....5H, 2021arXiv210408379C, Benisty_2021}, have identified a ubiquity of annular gaps and rings, as well as multiple instances of asymmetric arcs (crescents) and spiral arms, in the continuum emission. 

When applied to such high resolution observations, super-resolution techniques that fit the observed visibilities directly, such as \gal \citep{2018MNRAS.476.4527T} and \fr \citep{frank_method}, have found a yet greater occurrence rate of disc substructure. This includes identification of previously unseen features across the DSHARP survey \citep{frank_dsharp, 2021arXiv210508821A} and the ODISEA survey \citep{2021MNRAS.501.2934C}; in compact sources, including those that appear featureless in a \cl image \citep{2021A&A...645A.139K, 2021A&A...649A.122P}; and for observations at the highest available ALMA resolutions, such as in PDS~70 \citep{Benisty_2021}.

The next question is whether super-resolution techniques are also able to identify more substructure in moderate resolution observations. This would be particularly valuable for a statistical approach to substructure characterization over a large sample of discs, enabling a fuller investigation of demographic trends by exploiting the large archive of datasets at $\approx 100 - 300$~mas. 
This archive includes many discs that are not particularly large or bright, which current models predict should also contain substructure in order to counteract radial drift and retain reasonable dust disc sizes on few~Myr timescales \citep{2021arXiv210709914T}. 
We can ask for example whether compact discs that routinely appear smooth in \cl images are intrinsically featureless, or if this tends to be an artifact of observational or model resolution. 

\citet{Long2018} and \citet{Long2019} demonstrated at the survey level that parametric visibility fits can identify more substructure in moderate resolution ($120$~mas, $\approx 16$~au) observations than the \cl images alone.
Here we will push super-resolution visibility fits to still higher resolution, using the {\it nonparametric} approach in \fr to fit the observed visibilities yet more accurately. This will allow us to investigate how much more substructure in the Taurus survey data can be recovered from the observed visibilities -- including in compact sources -- and whether the identified features suggest new trends. 

In this work we characterize new substructure in $10$ of the Taurus survey discs using the 1D code \fr, which reconstructs a disc's brightness profile at super-resolution scales by nonparametrically fitting the azimuthally averaged visibility distribution.\footnote{The code is available at \href{https://discsim.github.io/frank}{\color{linkcolor}{https://discsim.github.io/frank}}.} Sec.~\ref{sec:model} summarizes the \fr modeling approach and its limitations. 
Sec.~\ref{sec:methodologies} more closely examines the major advantages (Sec.~\ref{sec:advantage}) and limitations (Sec.~\ref{sec:limitation_extrapolation} -- ~\ref{sec:limitation_asymmetry}) of nonparametric visibility fitting for datasets at the Taurus survey resolution, exploring how they affect substructure inference in \fr fits to these observations. 
In Sec.~\ref{sec:result} we present fits for the $10$ sources, grouping substructure findings into trends in compact discs (Sec.~\ref{sec:compact}) and extended discs (Sec.~\ref{sec:extended}). We further divide the extended sources into those with an inner and outer disc (Sec.~\ref{sec:extended_gaps}) and those with an inner cavity (Sec.~\ref{sec:extended_cavities}).
Sec.~\ref{sec:conclusion} summarizes our findings and briefly places them in the context of super-resolution substructure found in datasets outside the survey.

\section{Model}
\label{sec:model}
A full description of the \fr model framework and its limitations is in \citet{frank_method}. In short, \fr reconstructs the 1D (axisymmetric) brightness profile of a source as a function of disc radius by directly fitting the real component of the deprojected, unbinned visibilities as a function of baseline. The brightness profile is determined nonparametrically by fitting the visibilities with a Fourier-Bessel series, which is linked to the real space profile by a discrete Hankel transform. A Gaussian process regularizes the fit, with the covariance matrix nonparametrically learned from the visibilities under the assumption that this matrix is diagonal in Fourier space. The free parameters (diagonal elements) of the matrix correspond to the power spectrum of the reconstructed brightness profile. The fitting procedures takes $\lesssim 1$~min on a standard laptop for each dataset shown here.

To obtain the results shown in this work, we vary three of the five \fr model hyperparameters across datasets: $R_{\rm max}$, $N$ and \al. The hyperparameters $R_{\rm max}$ and $N$ simply set the maximum radius of the fit and number of brightness points in the fit, which we increase for larger discs. \al controls the prior on the Gaussian process, effectively determining
the signal-to-noise ({\it SNR}) threshold at which the model no longer attempts to fit the data. By varying \al we can thus account for the unique visibility distribution and noise properties of each dataset, with higher \al values imposing a stronger constraint that in practice causes the model to stop fitting the data at shorter maximum baseline.
Most of the Taurus survey datasets become noise-dominated at their longest baselines, as $(u,v)$ plane sampling becomes increasingly sparse. 
We will thus choose \al such that we fit the datasets out to long baselines, but stop before fitting clearly noise-dominated data (using $\alpha \in [1.01, 1.10]$). Pushing a fit out to long baselines to extract higher resolution information does nonetheless come at the cost of fitting some noise. The noise imprints on the brightness profile as short period, low amplitude oscillations; we will note nontrivial instances.

There are three major limitations in the current version of \fr:
\begin{enumerate}
\item A \fr fit drives to a visibility amplitude of zero once it stops fitting the data. This is intentional given the difficulty of generically extrapolating a fit beyond the edge of the observed visibilities, but we often expect the true visibility distribution would continue oscillating beyond the longest observed baselines if the disc is sufficiently structured.
We will motivate how this affects substructure inference in datasets characteristic of the Taurus survey in Sec.~\ref{sec:limitation_extrapolation}.
Ultimately this issue stems from the ill-posed nature of reconstructing the sky brightness from Fourier data, and it is also why the uncertainty on a \fr brightness profile is easily underestimated, particularly for deep gaps (\cl brightness profiles can similarly exhibit underestimated uncertainties for this reason). We thus will not show uncertainties in the \fr profiles in this work.
\item The 1D approach in \fr fits for the azimuthal average of the visibility data at each baseline. While this is an accurate representation of the azimuthally averaged brightness profile, in the presence of azimuthal asymmetries the brightness profile should be interpreted with caution, as (particularly super-resolution) asymmetries can be misidentified as annular features. We will demonstrate this in Sec.~\ref{sec:limitation_asymmetry}. 
\item The \fr real space model is not positive definite and so can exhibit regions of small amplitude, negative brightness. When this unphysical behavior occurs we can enforce positivity by finding the most probable brightness profile for a given set of power spectrum parameters and the constraint that the brightness be nonnegative, using a nonnegative least squares solver. This sometimes alters features across the disc (i.e., not just in regions of negative brightness) because the enforced positivity condition affects the visibility fit at long baselines. We will remove this limitation in a forthcoming work and version of the code by fitting in logarithmic brightness space, but for the current analysis we will show nonnegative fits for those \fr models that would otherwise exhibit regions of negative brightness; we will note which fits include this correction.
\end{enumerate}

\subsection{Data reduction}
\label{sec:data_reduction}
\begin{figure*}
	\includegraphics[width=\textwidth]{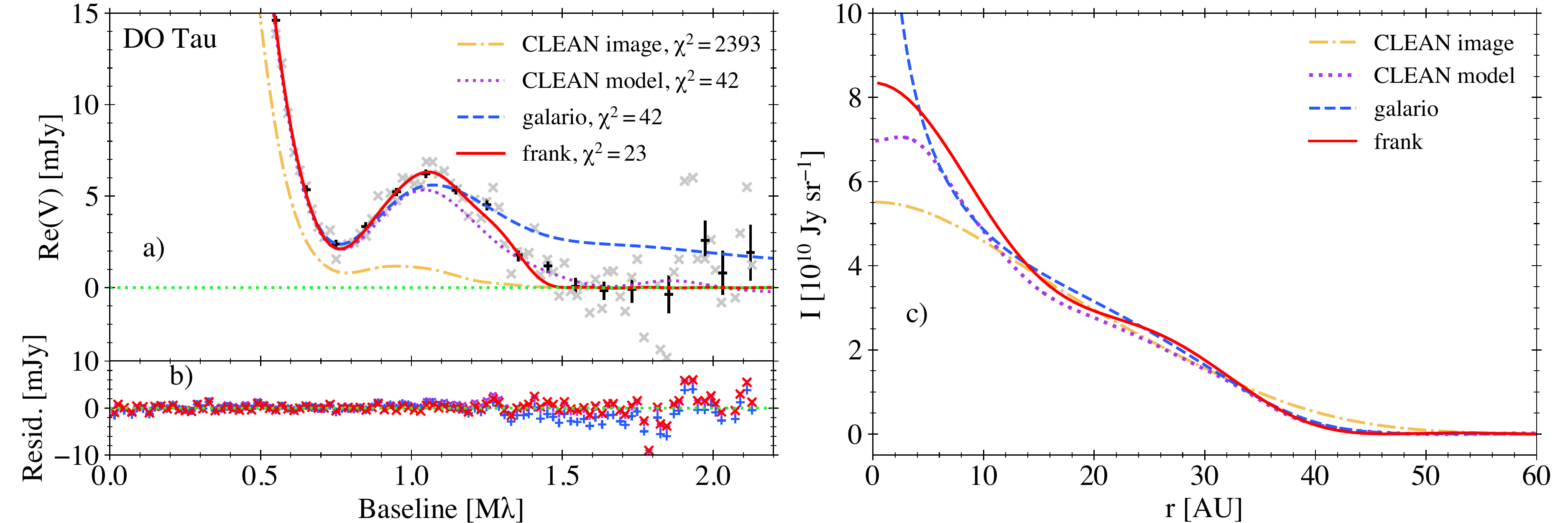}
	    \caption{{\bf Improved visibility fit accuracy better resolves disc structure} \newline
	    a) A zoom on the Taurus survey visibilities for DO~Tau ($20$ and $100$~\kl bins, with $1\sigma$ uncertainties shown for the $100$~\kl points); the parametric fit from \citet{Long2019}; the nonparametric \fr fit; and the Fourier transform of brightness profiles extracted from the \cl image and \cl model. \newline
	    b) Residuals of the parametric and \fr visibility fits and the \cl model transform ($20$~\kl bins). \newline
	    c) Brightness profiles for DO~Tau corresponding to the visibility fits in (a).
        }
    \label{fig:vis_accuracy}
\end{figure*}

In this work we reanalyze the ALMA Taurus survey published by \citet{Long2018} and \citet{Long2019}, to which we refer for details on the observational setup and calibration procedure.
To apply \fr to the datasets, we first apply channel averaging ($1$ channel per spectral window) and time averaging ($60$~s) to all spectral windows in the self-calibrated measurement set, then extract the unflagged visibilities.
We then use the disc geometries and phase centers in \citet{Long2019} to deproject the visibilities in \fr prior to fitting their 1D distribution. After deprojection, we re-estimate the weights by a constant factor of order unity to approximate the relation $w = 1 / \sigma^2$, where $w$ is the weight of a visibility point and $\sigma^2$ is the variance of its real and imaginary components.

\section{Methodologies -- Advantages and limitations of a 1D, nonparametric visibility fit}
\label{sec:methodologies}
Here we examine the benefits and drawbacks of 1D, nonparametric visibility fits (both generally and specific to \fr) for brightness profile reconstruction at resolutions typical of the Taurus survey, $\approx 120$~mas.

\subsection{Advantages -- A highly accurate fit to the observed data}
\label{sec:advantage}
\begin{figure*}
	\includegraphics[width=\textwidth]{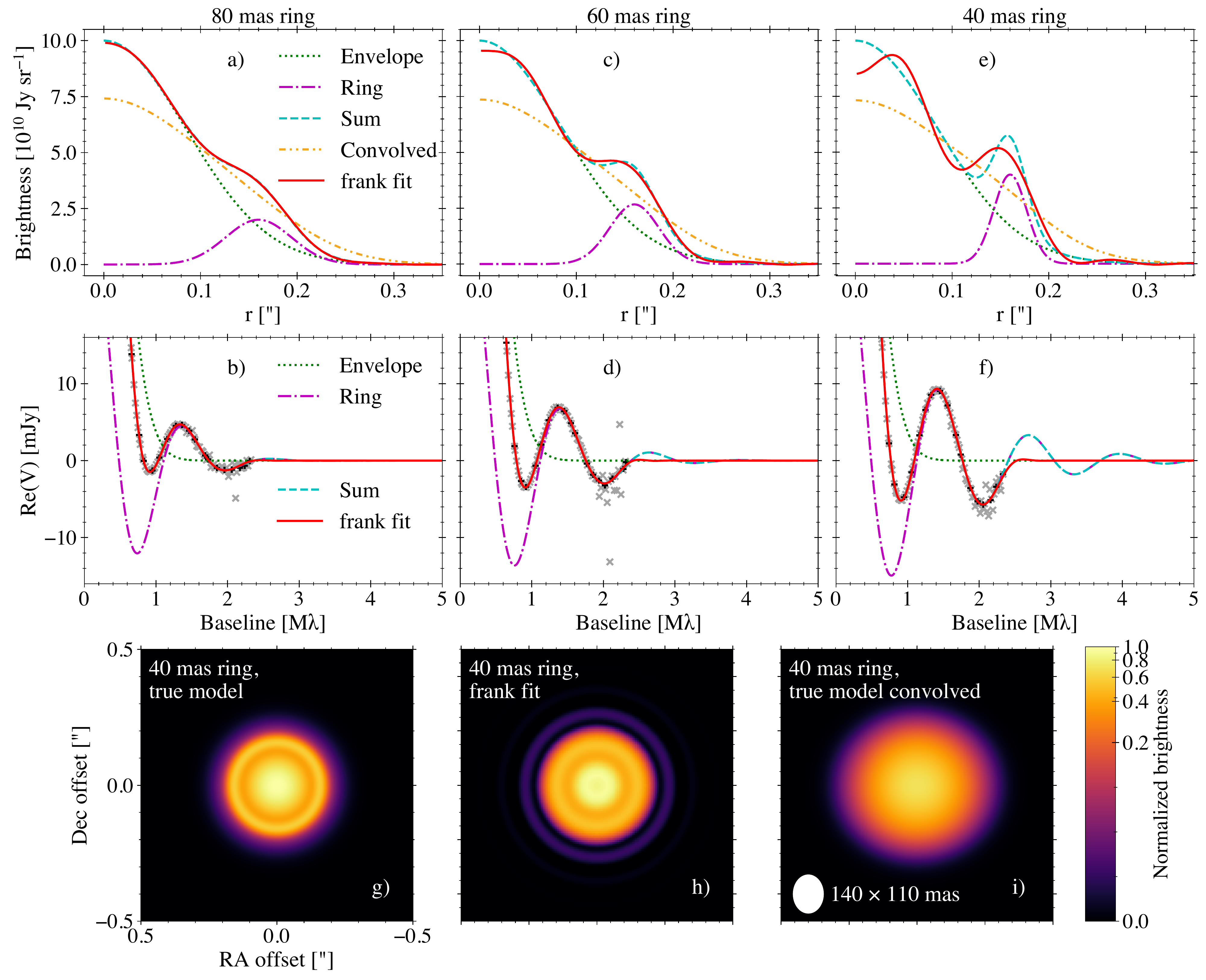}
	    \caption{{\bf \fr brightness profile accuracy decreases as a ring becomes increasingly super-resolution} \newline
	    a) Mock brightness profile of a compact disc with a shallow Gaussian ring; the profile is the sum of a Gaussian envelope and Gaussian ring, with each component shown. The ring's FWHM is given in the plot title. The \fr recovery of the summed profile is also shown, as is the summed profile convolved with a $140 \times 110$~mas beam. \newline
	    b) The real component of the 1D Fourier transform of the Gaussian envelope, ring, and their sum (the summed profile peaks at $\approx 150$~mJy). Also shown are noisy mock observations of the summed profile ($20$ and $100$~\kl bins), and the \fr fit to these mock data. \newline
	    c) -- d) and e) -- f) As in (a) -- (b), but with the Gaussian ring's FWHM successively decreased and amplitude correspondingly increased to conserve the disc's total (2D) flux. \newline
	    g) -- i) The noiseless, true model image of the disc in (e), the 1D \fr fitted profile swept over $360^{\rm o}$ in azimuth, and the true model convolved with the $140 \times 110$~mas beam. The images use an arcsinh stretch ($I_{\rm stretch} = {\rm arcsinh}(I/a)\ /\ {\rm arcsinh}(1/a),\ a = 0.02$) and the same absolute brightness normalization.
        }
    \label{fig:mock_fits}
\end{figure*}

Recovering super-resolution structure in a brightness profile with a 1D visibility model is a matter of fit accuracy; even a modest improvement in accuracy can correspond to new or more highly resolved profile features. 
To demonstrate how a nonparametric visibility fit's improved accuracy can better constrain super-resolution structure in Taurus survey data, {\bf Fig.~\ref{fig:vis_accuracy}} compares the parametric visibility fit from \citet{Long2019} for the compact disc DO~Tau with the nonparametric \fr fit.\footnote{All visibility fits from \citet{Long2018} and \citet{Long2019} shown in this work are obtained by taking the 1D Fourier transform of their best-fit \gal brightness profiles.}
\citet{Long2019} inferred that structure in the visibility distribution for this source indicates a sharp outer edge in the brightness profile, and so they modeled the profile parametrically as an exponentially tapered power law.
The resulting visibility fit in Fig.~\ref{fig:vis_accuracy}(a) is more accurate than the Fourier transform of a brightness profile extracted from the \cl image.\footnote{This difference is primarily due to the resolution loss induced by \cl beam convolution, which results in the transform of the \cl image poorly representing the observed visibilities. While we should thus not expect the transform of a \cl image profile to be accurate at long baselines, we will include this visibility profile in comparisons throughout this work because the \cl \texttt{.image} is the most common imaging product on which analysis is conducted in this field.} 
The parametric visibility fit's improved accuracy in turn corresponds to super-resolution structure recovery in the brightness profile; this structure is also apparent in a profile extracted from the \cl model.\footnote{All \cl brightness profiles for Taurus survey data in this work are extracted from \cl model images (the {\it .model} output of \texttt{tclean}) and convolved images (the {\it .image} output of \texttt{tclean}) generated using \texttt{tclean} in \texttt{CASA 5.6.1-8} with the \texttt{multiscale} deconvolver (pixel size of $30$~mas and scales of $1,\ 2,\ 4,\ 6$ pixels); a threshold of $3\sigma$, where $\sigma$ is the RMS noise measured in a region of the image far from the source; and Briggs weighting with a robust value of $0.5$.} 
While the parametric visibility fit is accurate at short and intermediate baselines, its residuals in Fig.~\ref{fig:vis_accuracy}(b) show nontrivial error at long baselines. By comparison, the \fr visibility model in Fig.~\ref{fig:vis_accuracy}(a) and its residuals in (b) demonstrate a yet higher accuracy across intermediate and long baselines.

We can quantify an improvement in fit accuracy with the $\chi^2$ statistic, \\
$\chi^2 = \sum\limits_{k=1}^{N} w_k [Re(V_{k,\ {\rm obs}}) - Re(V_{k,\ {\rm fit}})]^2$,
where we neglect the imaginary component of the visibilities because \fr only fits the real component.
As given in the legend of Fig.~\ref{fig:vis_accuracy}(a), both the parametric fit and the Fourier transform of a brightness profile extracted from the \cl model exhibit a smaller $\chi^2$ than the transform of a profile extracted from the \cl image by a factor of $57.0$ for this source, while the \fr fit yields a further reduction of the $\chi^2$ value by a factor of $1.8$. This comparatively small improvement in fit accuracy with \fr corresponds to a clear change in the disc morphology in the \fr brightness profile in panel (c), with the bump at $28$~au in the \fr profile not seen in the parametric profile and only hinted at in the \cl model profile. (Note that while the \cl model profile has lower integrated flux than the \fr profile -- because there is visibility information left in the residuals during the \cl process -- reducing the \texttt{tclean threshold} value also results in fitting more noise.)
Thus even in a dataset with a simple visibility distribution and relatively featureless brightness profile, a fairly small improvement to the accuracy of a visibility fit can nontrivially inform the scale and location of super-resolution structure in the recovered profile. This is the main advantage of a nonparametric fit, and it motivates why, for the sources in Sec.~\ref{sec:result} which all exhibit more structured visibility distributions than DO~Tau, a more accurate visibility fit with \fr yields new brightness profile features (as well as more highly resolved known features) relative to the parametric fits and the \cl models.

\subsection{Limitations -- Extrapolating the fit to unobserved baselines}
\label{sec:limitation_extrapolation}
\begin{figure*}
	\includegraphics[width=\textwidth]{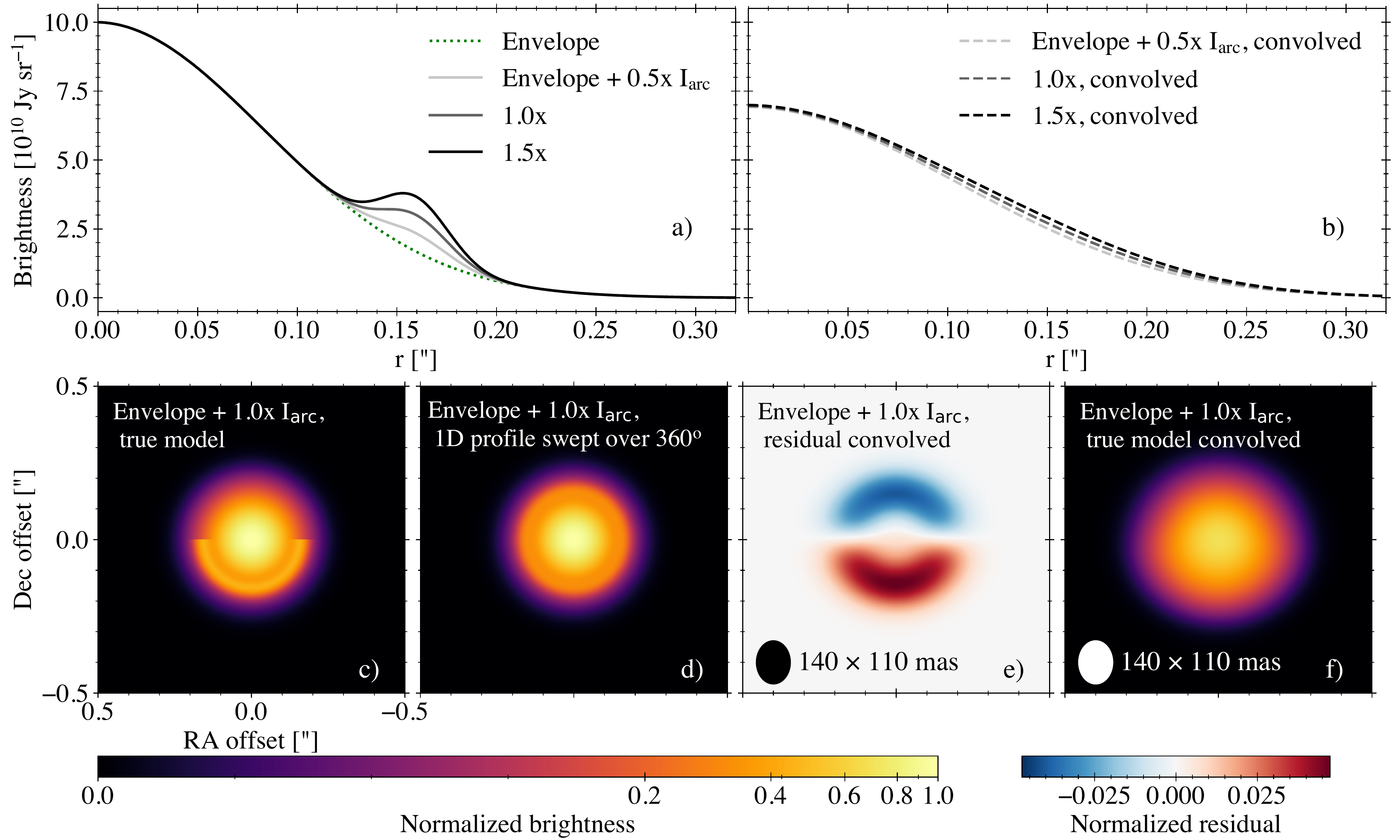}
	    \caption{{\bf Low contrast, asymmetric substructure emulates an underresolved annular ring in a brightness profile} \newline
	   a) The brightness profile of a Gaussian envelope, as well as the azimuthally averaged profile of the envelope summed with a $40$~mas arc that spans $180^{\rm o}$ in azimuth. The summed profile is shown for various amplitudes of the arc ($0.5$x, $1.0$x and $1.5$x an arbitrary value of $I_{\rm arc} = 2.67 \times 10^{10}$~Jy~sr$^{-1}$). \newline
	   b) Brightness profiles obtained from the 2D images of the envelope~$+$~arc, when convolved with a Gaussian beam whose $140 \times 110$~mas size is typical of the Taurus survey. \newline
	   c) -- f) For the $1.0$x $I_{\rm arc}$ case: the noiseless, true model image; the 1D brightness profile extracted from this image [which is shown in (a)] swept over $360^{\rm o}$; the residual between these two images, convolved with the beam; and the true model image convolved with the beam [corresponding to the brightness profile in (b)]. The disc images use an arcsinh stretch ($I_{\rm stretch} = {\rm arcsinh}(I/a)\ /\ {\rm arcsinh}(1/a),\ a = 0.02$) and the same absolute brightness normalization; the residual image uses a linear stretch symmetric about zero.
        }
    \label{fig:mock_asym}
\end{figure*}

A fair question to then ask is how much we trust the morphology of features in a super-resolution profile. An important consideration is that even super-resolution fits can still be expected to underresolve most disc features (even broad ones, albeit to a lesser extent), as is evident when comparing fits to lower and higher resolution observations of the same source (differing in resolution by a factor of say $3$). True features in a disc that are highly super-resolution (very roughly, a factor $\gtrsim 3$ narrower than the nominal spatial resolution) tend to be inaccurately recovered in a \fr fit, and in some cases they can induce erroneous oscillations in the brightness profile, as we will now show.

{\bf Fig.~\ref{fig:mock_fits}} demonstrates the accuracy of a \fr fit to a disc with super-resolution features using mock data. In Fig.~\ref{fig:mock_fits}(a) -- (b), we first consider a simple disc -- the sum of a Gaussian envelope and a shallow Gaussian ring whose $80$~mas full width at half maximum ({\it FWHM}) is super-resolution relative to the $\approx 120$~mas FWHM beam of the mock observations by a factor of $\approx 1.5$.\footnote{The mock dataset is generated with a baseline distribution and noise properties that emulate the Taurus survey observations of DR~Tau.} 
While the profile convolved with a $140 \times  110$~mas beam (typical of the Taurus survey) in (a) shows no clear indication of the super-resolution ring, fitting the visibilities in (b) with \fr gives an accurate recovery of the true brightness profile. 
But if we then narrow the ring to $60$~mas (and increase its surface brightness to conserve total flux) in Fig.~\ref{fig:mock_fits}(c) -- (d), it is now super-resolution by a factor of $\approx 2$, and the \fr recovered profile begins to show some clear inaccuracy. It exhibits a plateau around $0.16 \arcsec$, underresolving the true gap/ring pair. This is due to an inaccurate extrapolation of the \fr visibility fit beyond the mock observation's longest baselines, where the true profile's visibility distribution continues to oscillate.
A further consequence of the fit's underestimated visibility amplitudes at unsampled baselines is the underestimated peak brightness in the \fr brightness profile.

Narrowing and brightening the ring even further so that it has a $40$~mas FWHM (super-resolution by a factor of $\approx 3$) in Fig.~\ref{fig:mock_fits}(e) -- (f), the convolved profile in panel (e) [and the 2D image of this profile swept over $2\pi$ in panel (i)] still shows no hint of the ring. 
The \fr profile in (e) identifies the gap/ring pair, but underresolves the feature amplitudes and misidentifies their centroids. The \fr profile also underestimates the peak brightness more severely, showing an erroneous turnover near $r=0$. This turnover is a consequence of the narrower ring in the true profile increasing the absolute visibility amplitudes at all baselines; accurately fitting the higher amplitude features in the visibilities introduces higher contrast structure into the brightness profile. Because the \fr fit has a visibility amplitude of $\approx 0$ beyond the edge of the data, while the true visibility distribution has nontrivial amplitude there, these higher contrast structures are not well constrained. This effect also introduces the erroneous, shallow bump into the \fr brightness profile between $0.2 - 0.3 \arcsec$, appearing in the 2D image of the swept \fr brightness profile in Fig.~\ref{fig:mock_fits}(h) as a faint but fake ring [compare the true 2D image in panel (g)].
It is thus possible for highly super-resolution features in a true brightness profile to introduce erroneous oscillations into a \fr brightness profile.

For some datasets in the Taurus survey such as DO~Tau in Fig.~\ref{fig:vis_accuracy}(a), this is not much of a concern, as the observed visibilities appear to plateau at zero at the longest baselines. But for other datasets it is less clear whether higher resolution and/or deeper observations would show the visibilities to continue oscillating beyond the baselines at which the current data become noise-dominated. 
While any extrapolation of a fit beyond the data's longest baselines is highly uncertain, it can be useful to compare a \fr brightness profile to that obtained with a parametric visibility fit, where the parametric profile's functional form is motivated for example by the \fr fit or by structure in the observed visibilities (as \citealt{Long2018} and \citealt{Long2019} have done). We will perform an in-depth comparison in Sec.~\ref{sec:extended_gaps} for the most structured disc in our results, DL~Tau. 

\subsection{Limitations -- Distinguishing azimuthally symmetric from asymmetric substructure}
\label{sec:limitation_asymmetry}
If we have found a super-resolution feature in a disc, the next question is whether it is an annular ring (gap) or an azimuthally asymmetric brightness excess (depletion). 
Because a 1D brightness profile averages the flux in a given annulus over $2\pi$ in azimuth, a low -- moderate contrast asymmetric feature within that annulus can mimic an underresolved (or shallow) ring in the profile. {\bf Fig.~\ref{fig:mock_asym}} demonstrates this with mock data, using a Gaussian disc with an additional brightness \lq{}arc\rq{} that is produced by sweeping a Gaussian ring only over $180^{\rm o}$ in azimuth in panel (c). The arc emulates a brightness excess on top of the background envelope, and a 1D profile in Fig.~\ref{fig:mock_asym}(a) extracted from the image in (c) shows a slight bump at the arc's radial location. From the brightness profile alone this could be misidentified as an annular feature, and because the arc is super-resolution by a factor of $\approx 3$, the true model image convolved with a $140 \times 110$~mas beam in (f) -- and the corresponding convolved brightness profile in (b) -- show no clear indication of it. When we increase the asymmetry's brightness by $50\%$, it emulates a shallow gap/ring pair in (a), while the convolved profile in (b) is effectively unchanged.

How then can we distinguish super-resolution asymmetries from annular features? We do not have an unambiguous method for this, so jointly consider three metrics: contouring the \cl image, identifying structure in the imaginary component of the visibilities, and imaging a \fr fit's residual visibilities. 
The first of these, contouring the \cl image at levels of the RMS noise, can be useful in identifying the convolved representation of super-resolution asymmetries. A limitation is that low contrast or sufficiently narrow features are often not identifiable.
Second, while an asymmetric feature is represented in the real component of the 1D visibilities exactly as an annular feature at the same location and that has the same width and total surface brightness (as integrated over $360^{\rm o}$ in azimuth),\footnote{We can intuit this by recalling that the Fourier transform is a linear operation; the Fourier transform of a feature is equal to the sum of its' components' Fourier transforms. Thus the transform of a ring is equal to the sum of the transforms of its azimuthal segments.} structure in the imaginary component of the visibilities indicates scales at which there is asymmetry with respect to the phase center. A limitation here is that without a robust model to fit Im($V$), interpretation of its structure can be complicated by the comparatively low amplitude (and thus low binned SNR) relative to Re($V$), and by the typical uncertainty in the disc phase center of $\lesssim 3$~mas. 
Third, imaging the \fr residual visibilities effectively isolates azimuthal asymmetries in the image by subtracting out the (fitted) average brightness at each radius. A limitation is that there is typically ambiguity in interpreting structure in imaged residuals, due to potential artifacts of an incorrect disc geometry and/or phase center, imaging artifacts, and loss in resolution by convolving the residuals with the \cl beam [as demonstrated in Fig.~\ref{fig:mock_asym}(e)].
While each of these three approaches is thus imperfect, together they can aid in distinguishing super-resolution asymmetries from annular features.

\section{Results \& Analysis}
\label{sec:result}
\begin{figure*}
	\includegraphics[width=\textwidth]{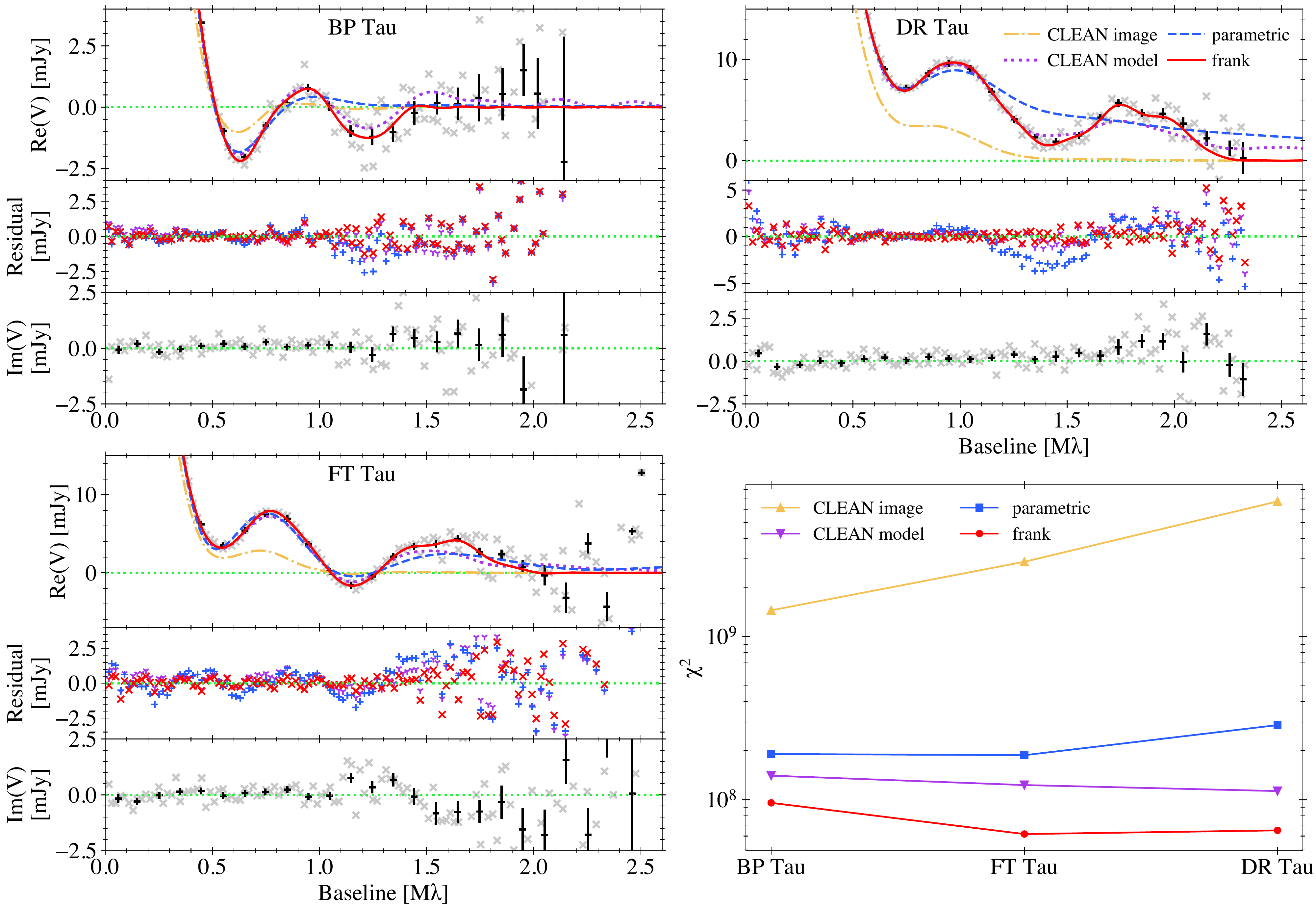}
	    \caption{{\bf Improved visibility model accuracy in \fr fits to compact discs} \newline
	    For the compact discs in Sec.~\ref{sec:compact}, a zoom on the visibilities ($> 0.30$~\ml; $20$ and $100$~\kl bins, with $1 \sigma$ uncertainties shown for the $100$~\kl points). The parametric visibility fit from either \citet{Long2018} or \citet{Long2019}, the \fr fit, and the Fourier transforms of the \cl image and model brightness profiles are shown. Also shown are residuals for the parametric and \fr fits and the \cl model transform ($20$~\kl bins; larger amplitude residuals at the longest baselines are beyond the y-range in some panels), as well as the imaginary component of the observed visibilities. Discs are arranged from top to bottom and then left to right in increasing \fr fit resolution. The bottom-right panel shows the $\chi^2$ statistic for each fit. 
        }
    \label{fig:vis_fits_compact}
\end{figure*}

\begin{figure*}
	\includegraphics[width=\textwidth]{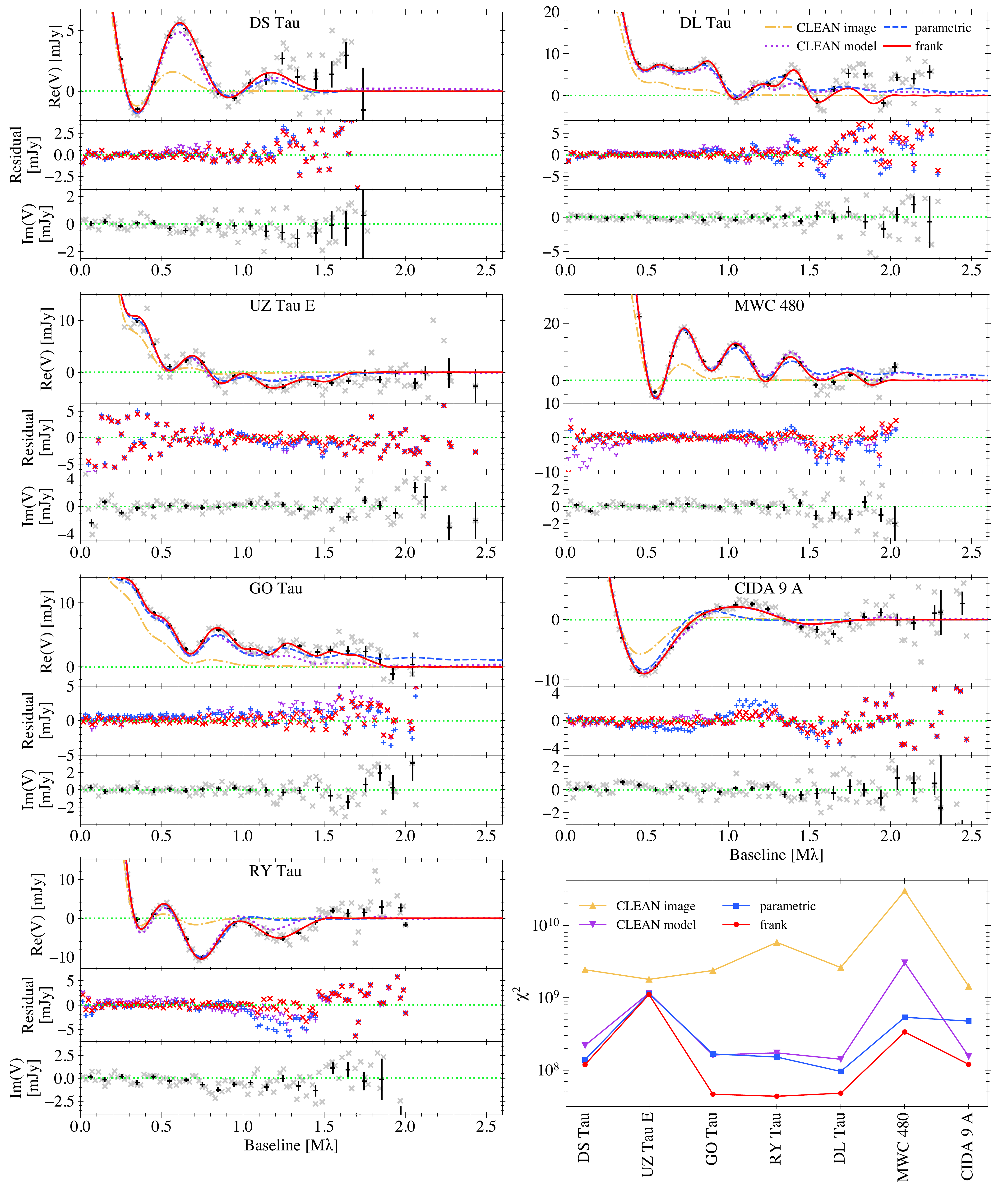}
	    \caption{{\bf Improved visibility model accuracy in \fr fits to extended discs} \newline
	    \ As in Fig.~\ref{fig:vis_fits_compact}, but for the extended discs in Sec.~\ref{sec:extended}.
        }
    \label{fig:vis_fits_extended}
\end{figure*}

\begin{figure*}
	\includegraphics[width=\textwidth]{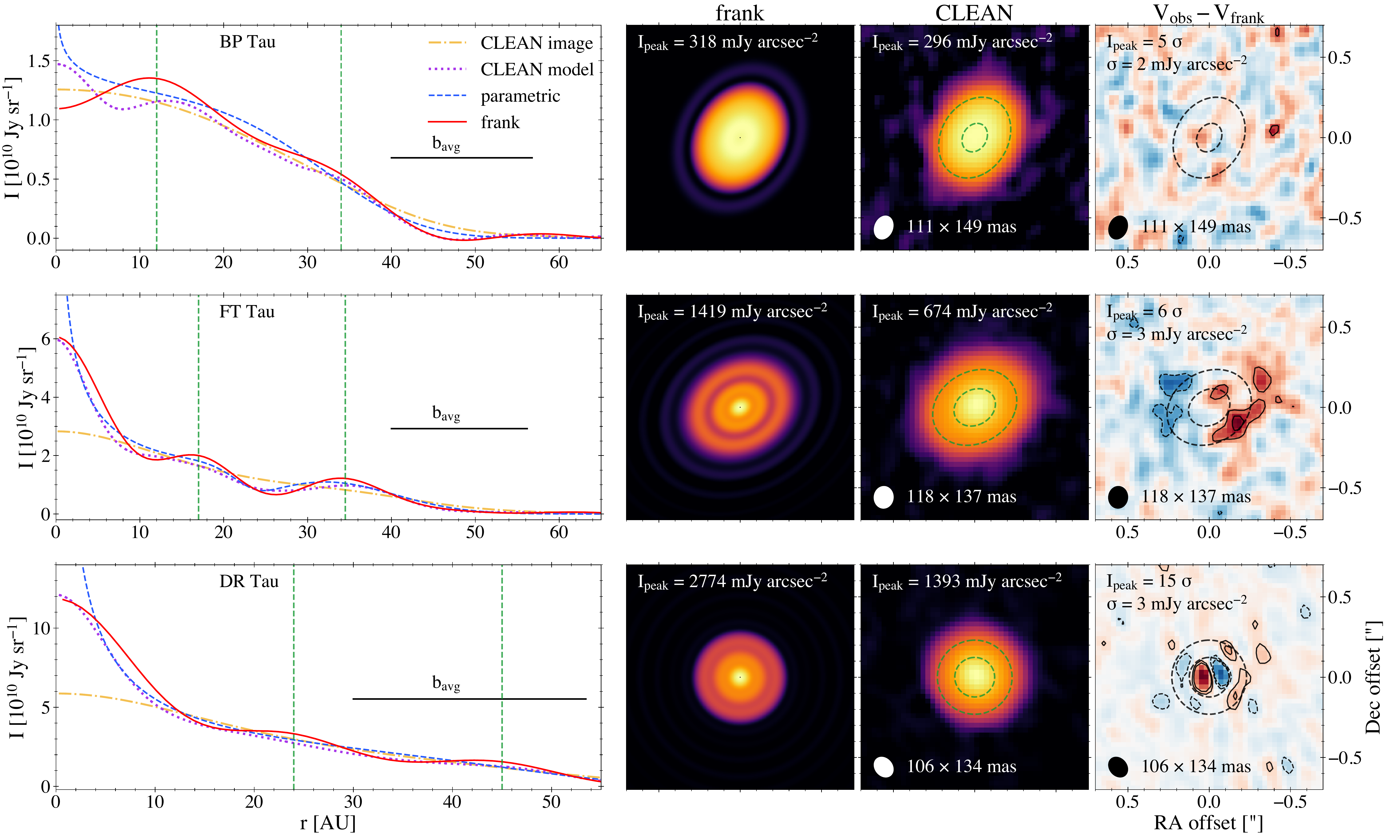}
	    \caption{{\bf New substructure in \fr fits to compact discs} \newline
	    Brightness profiles extracted from the \cl image and model, the parametric profile and \fr profile for three of the compact ($R_{\rm eff,\ 90\%} \leq 50$~au) systems in the Taurus survey, corresponding to the visibility fits in Fig.~\ref{fig:vis_fits_compact} (the parametric fits are from \citealt{Long2018} and \citealt{Long2019}; $b_{\rm avg}$ shows the mean of the \cl beam width along its major and minor axes). Also shown are an image of the \fr profile swept over $2\pi$ and reprojected, the \cl image, and the imaged \fr residual visibilities (zero \cl iterations; contours at $-5, -3, +3, +5 \sigma$). Vertical lines in the brightness profile plots denote features that are shown as ellipses in the \cl image and imaged \fr residuals for reference. The \fr and \cl images use an arcsinh stretch ($I_{\rm stretch} = {\rm arcsinh}(I/a)\ /\ {\rm arcsinh}(1/a),\ a = 0.02$), but different brightness normalization (indicated by the given peak brightness). The imaged \fr residuals use a linear stretch symmetric about zero. We use image brightness units of [mJy arcsec$^{-2}$] to facilitate comparison between datasets of different beam size.
        }
    \label{fig:compact_discs}
\end{figure*}

\begin{figure*}
	\includegraphics[width=\textwidth]{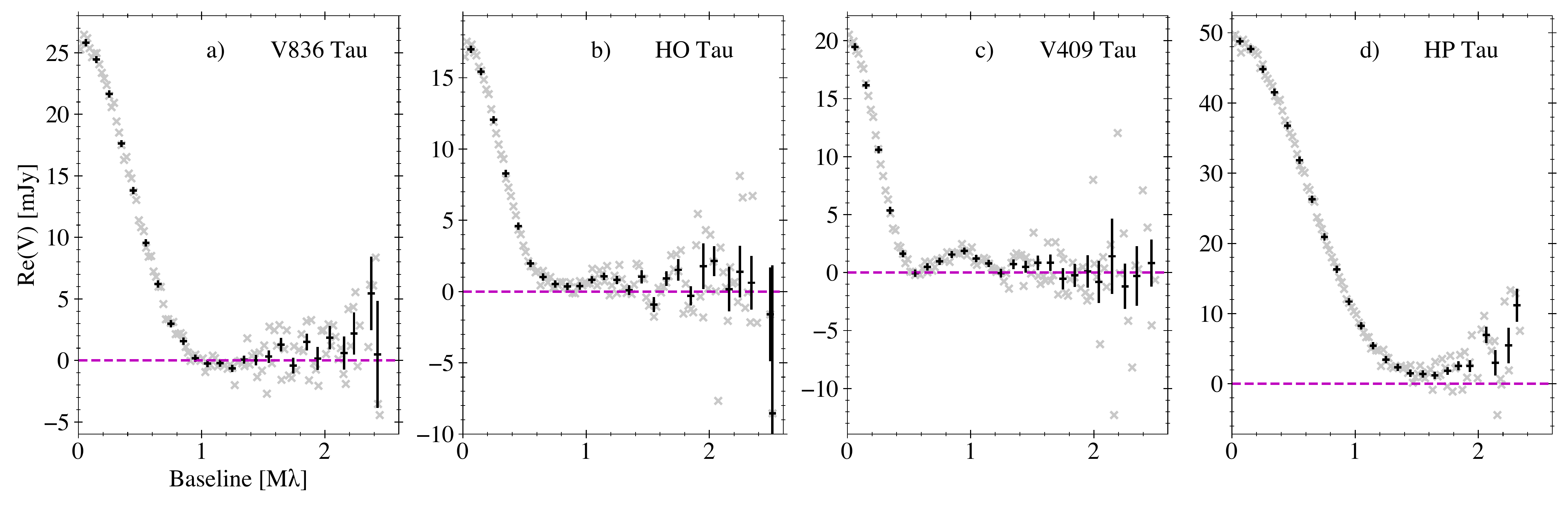}
	    \caption{{\bf Noisy visibility distributions for compact discs} \newline
	    Visibility distributions for four compact ($R_{\rm eff,\ 90\%} \leq 50$~au) sources in the Taurus survey, whose structure at long baselines is unclear due to $(u, v)$ plane sparsity.
        }
    \label{fig:compact_vis_additional}
\end{figure*}

Of the $32$ sources in the Taurus survey, our analysis focused only on the $24$ single-disc systems; among these, here we show the $10$ for which we obtain a brightness profile with prominent new substructure. The remaining $14$ comprise four extended discs where the \fr visibility fit is highly similar to the parametric fit in \citet{Long2018} and $10$ compact discs whose fitted brightness profiles lack substructure. In Sec.~\ref{sec:compact} we will discuss the general potential for substructure in these compact sources based on the observed visibility distributions. All \fr fits in this work are available at \href{https://zenodo.org/record/6686456}{\color{linkcolor}{https://zenodo.org/record/6686456}}.

The \fr fit hyperparameters for the $10$ datasets where we find new substructure are summarized in {\bf Table~\ref{tab:frank_pars}}. 
We divide our analysis into compact and extended discs. The compact discs -- BP~Tau, DR~Tau and FT~Tau -- have an effective radius $R_{\rm eff,\ 90\%} \leq 50$~au, where the integrated flux $f(R_{\rm eff,\ 90\%}) = 0.9 \cdot 2 \pi \int_0^{r=\infty} I(r)\ r\ dr$. The extended discs -- CIDA~9~A, DL~Tau, DS~Tau, GO~Tau, MWC~480, RY~Tau, and UZ~Tau~E -- have $R_{\rm eff,\ 90\%} > 60$~au. 
\citet{Long2019} fit BP~Tau and DR~Tau parametrically with \gal, using an exponentially tapered power law to model the brightness profile, motivated by structure in the observed visibility distributions. \citet{Long2018} fit the remaining eight discs shown here with a parametric form in \gal comprised of a sum of Gaussians (for CIDA~9~A, DS~Tau, RY~Tau, and UZ~Tau) or an exponentially tapered power law summed with Gaussians (for DL~Tau, FT~Tau, GO~Tau, and MWC~480). Their choice of the number of Gaussians for each source is motivated by a brightness profile extracted (along the disc's major axis) from the \cl image.

\begin{table}
\caption{For each Taurus survey disc in Sec.~\ref{sec:methodologies} and~\ref{sec:result}, the distance to the source (using Gaia DR2 measurements from \citealt{Bailer-Jones:2018aa}),
and the values for the five \fr hyperparameters: SNR criterion \al, strength of smoothing \ws applied to the reconstructed power spectrum, outer radius of the fit $R_{\rm out}$, number of radial (and spatial frequency) points $N$, and floor value $p_0$ for the reconstructed power spectral mode amplitudes. Sec.~\ref{sec:model} gives a fuller explanation of \al. All \fr fits in this work are available at \href{https://zenodo.org/record/6686456}{\color{linkcolor}{https://zenodo.org/record/6686456}}.}
\begin{tabular}{l c c c c c c}
    \hline
    {\bf Disc} & $\boldsymbol{d}$ {\bf [pc]} & $\boldsymbol{\alpha}$ & $\boldsymbol{\log_{10}}$ & $\boldsymbol{R_{\rm out}\ [\arcsec]}$ & $\boldsymbol{N}$ & $\boldsymbol{p_0}$ \\
    & & & $\boldsymbol{w_{\rm smooth}}$ & & & $\boldsymbol{{\rm [Jy^2]}}$ \\
    \hline
    \multicolumn{7}{c}{Compact discs} \\
    \hline
    BP~Tau$^\dagger$ & 129 & 1.01 & -4 & 1.0 & 200 & $10^{-15}$ \\    
    DO~Tau & 139 & " & " & " & " & " \\  
    DR~Tau & 195 & " & " & " & " & " \\  
    FT~Tau & 127 & " & " & " & " & " \\ 
    \hline
    \multicolumn{7}{c}{Extended discs} \\
    \hline
    UZ~Tau~E & 131 & 1.01 & -4 & 1.0 & 200 & $10^{-15}$ \\
    CIDA~9~A$^\dagger$ & 171 & 1.05 & " & " & " & " \\  
    DS~Tau$^\dagger$ & 159 & 1.05 & " & " & " & " \\ 
    RY~Tau & 128 & 1.10 & " & " & " & " \\
    DL~Tau$^\dagger$ & 159 & 1.01 & " & 1.5 & 300 & " \\     
    GO~Tau & 144 & 1.00 & " & 1.5 & 300 & " \\             
    MWC~480$^\dagger$ & 161 & 1.01 & " & 1.5 & 300 & " \\         
    \hline
    \multicolumn{7}{l}{$^\dagger$ These fits have enforced brightness profile positivity (see Sec.~\ref{sec:model}).}
\end{tabular}
\label{tab:frank_pars} 
\end{table}

For each of these $10$ sources, we compare the \fr visibility fit to the parametric fit, as well as the Fourier transform of the \cl brightness profile, in {\bf Fig.~\ref{fig:vis_fits_compact}} for the compact discs (the corresponding brightness profiles, discussed below, are in Fig.~\ref{fig:compact_discs}) and {\bf Fig.~\ref{fig:vis_fits_extended}} for the extended discs (brightness profiles in Figs.~\ref{fig:extended_discs_rings} and~\ref{fig:extended_discs_cavities}).
In every case the parametric model matches the data more accurately than the Fourier transform of the \cl image profile, the transform of the \cl model profile is generally comparable to or in some cases slightly more accurate than the parametric model, and the residuals and $\chi^2$ values demonstrate that the \fr fit is more accurate than each of the \cl image, \cl model and parametric model visibility profiles. The \fr fits that have enforced brightness profile positivity (BP~Tau, CIDA~9~A, DL~Tau, DS~Tau, MWC~480) underestimate data amplitudes at long baselines. Nonetheless, we recall from Sec.~\ref{sec:advantage} that even modest improvements in visibility fit accuracy can yield appreciably more highly resolved brightness profile features, and in some cases can identify new features. 

In order to examine whether new features may be nonaxisymmetric, Fig.~\ref{fig:vis_fits_compact} and~Fig.~\ref{fig:vis_fits_extended} also show the imaginary component of the observed visibility distributions (which \fr treats as zero at all baselines). We will discuss Im($V$) in relation to disc asymmetries in the following subsections. 

\subsection{New substructure in compact discs}
\label{sec:compact}
Across three compact sources in the survey -- BP~Tau, DR~Tau and FT~Tau -- the \fr fits in {\bf Fig.~\ref{fig:compact_discs}} find new substructure. Additionally we note that for the highly compact ($R_{\rm eff,\ 90\%} < 25$~au) disc T~Tau~N, in which \citet{2021ApJ...923..121Y} recently found a gap/ring pair with their 2D, super-resolution modeling framework \texttt{PRIISM}, the \fr fit (not shown here) demonstrates agreement in the gap/ring pair's location and approximate amplitude. \newline

\noindent {\bf BP~Tau:} The \fr fit to BP~Tau identifies a new turnover in the inner disc that is not seen in the \citet{Long2019} parametric profile because the corresponding visibility fit does not recover the negative peak in the data at $1.25$~\ml.
The parametric model instead finds an almost flat inner disc (power law index of $0.1$), resulting in a quasi-linear region of the brightness profile between $\approx 8 - 17$~au; this can be understood as a result of underresolving the turnover (which may itself be an underresolved ring). 
Representation of an underresolved brightness excess as a quasi-linear region in a brightness profile is demonstrated with mock data in Fig.~\ref{fig:mock_asym}(b).
We can further motivate the turnover by the observed visibilities; their amplitudes are preferentially negative between $\approx 0.5 - 1.5$~\ml, which is an indication of a wide Gaussian in the brightness profile that is not centered at zero radius. Notice how the visibility distributions in Fig.~\ref{fig:vis_fits_extended} for the three discs with an apparent inner cavity (and thus, a Gaussian ring not centered at zero) -- CIDA~9~A, RY~Tau and UZ~Tau~E -- also each exhibit preferentially negative visibility amplitudes at intermediate baselines. The \fr profile also better localizes the structure beyond $20$~au in the disc than the parametric profile, with the \cl model profile showing rough agreement with \fr here.

To further assess the \fr profile features (using BP~Tau as an example for the analysis we will more succinctly cover in subsequent discs), we can consider how the model limitations in Sec.~\ref{sec:limitation_extrapolation} and~\ref{sec:limitation_asymmetry} may affect the fit.
Given the demonstration in Sec.~\ref{sec:limitation_extrapolation} of the difficulty in accurately extrapolating a fit to unobserved (and noise-dominated) baselines -- and how this can introduce fake oscillations into a profile when the underlying disc has highly super-resolution features -- we emphasize that a more accurate visibility fit or higher resolution/deeper observations can find the features in a \fr brightness profile to become either more or less prominent. This is particularly true in the innermost disc, where substructure can routinely be highly underresolved. The turnover in the \fr profile for BP~Tau for example may resolve into a ring or something more complicated, as may be indicated by the inner $15$~au of the \cl model profile in Fig.~\ref{fig:compact_discs}. We do expect the turnover to be indicating the presence of real substructure, given the dataset's preferentially negative visibility amplitudes at intermediate baselines as discussed above. In the outer disc, the broad, shallow feature in the \fr brightness profile between $\approx 52 - 65$~au is at least partly due to noise (influenced by the visibility fit's extrapolation of zero amplitude beyond $\approx 1.5$~\ml, analogous to Fig.~\ref{fig:mock_fits}(e)), but it may also have contributions from real, diffuse emission.

In light of the discussion in Sec.~\ref{sec:limitation_asymmetry} on how nonaxisymmetric features can mimic the appearance of a partially resolved ring, we can also use the \cl image, imaginary component of the visibilities and imaged \fr residuals to examine whether any super-resolution features in the \fr profile may be artifacts of azimuthally asymmetric emission. Contouring the \cl image of the source shows no clear signs of an asymmetry; the imaginary component of the visibility distribution in Fig.~\ref{fig:vis_fits_compact} does not exhibit prominent structure, indicating that asymmetries in the image must be particularly faint and/or small-scale; and the imaged \fr residuals do not have clear features within the disc (the small, $5\sigma$ blob in the west of the imaged residuals that is also in the \cl image). We thus infer that the profile's features are likely annular. \newline

\noindent {\bf FT~Tau:} The \fr fit in Fig.~\ref{fig:compact_discs} finds a new gap/ring pair around $11 - 17$~au, underresolved in the parametric brightness profile as the quasi-linear region (and hinted at in the \cl model profile). The \fr profile also determines the gap at $26$~au identified in \citet{Long2018} to be wider, with a brighter adjacent ring. 
While the difference between the parametric and \fr visibility fits for FT~Tau in Fig.~\ref{fig:vis_fits_compact} may not look dramatic enough to correspond to a new gap/ring pair, it is important first that \fr exhibits an improved fit accuracy over a large span in baseline ($\approx 1.0 - 1.7$~\ml). Second, while the \fr fit converges on zero visibility amplitude at $\approx 2.0$~\ml, the parametric fit remains positive and continues to slowly oscillate out to the longest baselines and beyond. The data instead appear by eye to indicate that the true visibility distribution becomes negative beyond $2.0$~\ml [denser $(u, v)$ plane sampling at these baselines would be needed to confirm].

Considering disc asymmetries, the imaginary component of the visibilities for FT~Tau in Fig.~\ref{fig:vis_fits_compact} have clear structure on scales between $\approx 1.2 - 1.7$~\ml, and this structure has amplitude comparable to the difference between the \fr and parametric fit residuals for Re($V$). The imaged \fr residuals also have $\leq 5 \sigma$ features within and beyond the gap at $\approx 26$~au. Together this suggests that there may be some faint asymmetric structure in the disc, particularly in the gap centered at $26$~au, where the residual amplitude is largest. \newline

\noindent {\bf DR~Tau:} We find two new gaps relative to \citealt{Long2019} (the \fr fit to DR~Tau was previously shown in \citealt{frank_method}) in Fig.~\ref{fig:compact_discs}. This can be motivated by the significant difference in visibility fit accuracy between the parametric and \fr models in Fig.~\ref{fig:vis_fits_compact}; the \cl model visibility profile is also more accurate than the parametric model, with the \cl model brightness profile having a hint of the outer ring found in the \fr brightness profile.  The qualitative similarity in structure between the observed visibility distributions for FT~Tau and DR~Tau also motivates why the \fr fit shows two gaps in both discs. The visibilities for DR~Tau do not exhibit a zero-crossing, indicating the data contain underresolved structure at small spatial scales; this seems most likely to be an indication of a partially resolved inner disc.

Considering the inner disc, while the imaginary component of the visibilities for DR~Tau do not show clear structure, the imaged \fr residuals in Fig.~\ref{fig:compact_discs} do have strong features in the innermost radii ($\leq 15 \sigma$, or $\lesssim 5\%$ of the background brightness in the \cl image). This is likely affecting the morphology of the inner gap in the \fr profile to some extent. 
We find that the inner disc residuals in DR~Tau and other discs discussed below are not attributable solely to an incorrect determination of the disc phase center (assessed by varying the applied phase center in {\bf Appendix~\ref{sec:appendix_phase_center}}). 

\subsubsection{Occurrence rate of substructure in compact discs}
The \fr fits to BP~Tau and DR~Tau raise the number of compact, single-disc systems with substructure from two -- FT~Tau and the cavity disc IP~Tau (shown in \citealt{Long2018}) -- to four, out of $14$ total in the survey. 
Among the $14$ compact discs, these four are neither the largest nor brightest, which prompts the question of whether more of the survey's compact objects may be structured. To partially address this, we can consider whether the survey data strongly exclude the presence of substructure in the remaining $10$ compact sources. 
The visibility distributions for four additional compacts discs in {\bf Fig.~\ref{fig:compact_vis_additional}} each show tentative or clear indications of structure at intermediate baselines and become highly noisy at longer baselines due to sparse $(u, v)$ plane sampling. Whether this structure at intermediate baselines corresponds in each case only to the brightness profile becoming steeper in the outer disc (i.e., no gap/ring substructure), or instead to substructure, is not clear from these data. The visibility distributions in Fig.~\ref{fig:compact_vis_additional}(b) -- (d) also do not exhibit a zero-crossing at short baseline, characteristic of an underresolved inner disc and/or highly super-resolution substructure. It is thus possible that higher resolution and/or deeper observations would identify substructure in a larger subset of the survey's compact sources.

That the current data do show substructure in four of the survey's compact discs -- BP~Tau, DR~Tau, FT~Tau, and IP~Tau -- is in line with multiple features detected in the \fr fits to the DSHARP observations of the compact sources SR~4, DoAr~33 and WSB~52 \citep{frank_dsharp}, as well as substructure recovered in the parametric visibility fits to the compact discs CIDA~1, MHO~6 and J0433 \citep{2021A&A...649A.122P, 2021A&A...645A.139K}. Collectively these results demonstrate that many compact discs are not intrinsically featureless; their lack of apparent substructure is instead in some cases an artifact of either the resolving power of the model applied to the data or of the data itself. 
This may be a tentative indication that a nontrivial fraction of compact dust discs follow the same evolutionary pathway as extended discs, which tend to be structured.

\subsection{New substructure in extended discs}
\label{sec:extended}
\begin{figure*}
	\includegraphics[width=\textwidth]{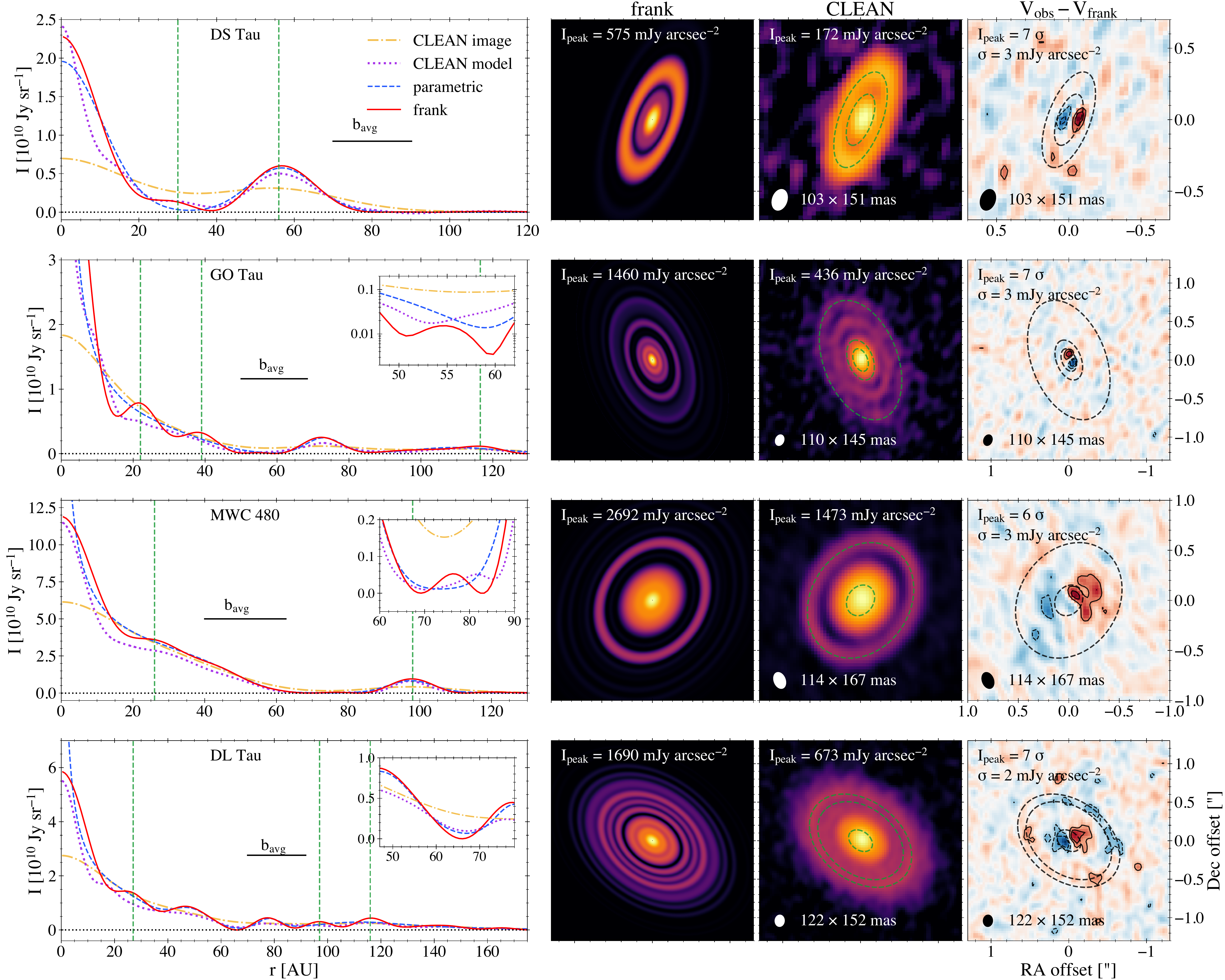}
	    \caption{{\bf New substructure in \fr fits to extended discs with outer rings} \newline
	    As in Fig.~\ref{fig:compact_discs}, but for the four extended ($R_{\rm eff,\ 90\%} > 50$~au) systems in the Taurus survey that exhibit an inner disc and one or more outer rings, discussed in Sec.~\ref{sec:extended_gaps}. Parametric profiles are from \citealt{Long2018}; the visibilities and fits for these discs are in Fig.~\ref{fig:vis_fits_extended}. 
	    The inset panels zoom on deep gaps in the brightness profiles. The \fr fit to GO~Tau peaks at $6.2 \times 10^{10}$~Jy~sr$^{-1}$.
        }
    \label{fig:extended_discs_rings}
\end{figure*}

\begin{figure*}
	\includegraphics[width=\textwidth]{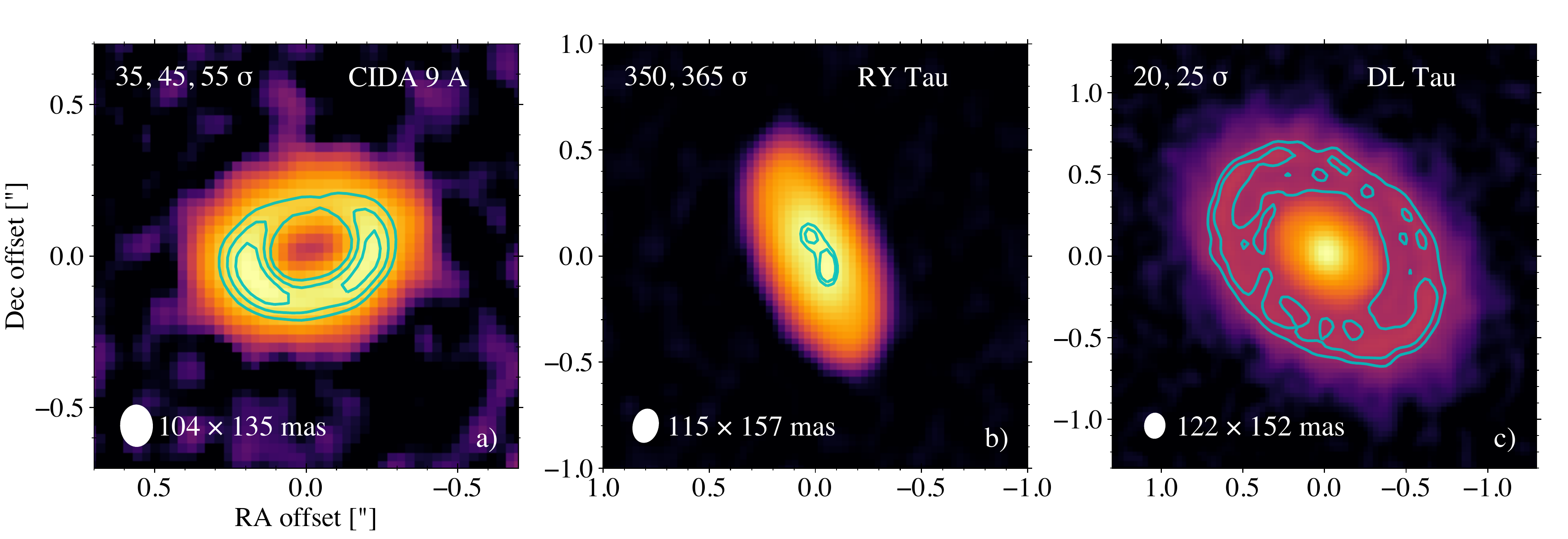}
	    \caption{{\bf \cl image asymmetries} \newline
	    \ \cl images for three of the Taurus survey's extended discs (see Sec.~\ref{sec:extended}), with contours chosen to highlight asymmetries. The images are identical to those in Fig.~\ref{fig:extended_discs_rings} and~\ref{fig:extended_discs_cavities}.
        }
    \label{fig:clean_contour}
\end{figure*}

\begin{figure*}
	\includegraphics[width=\textwidth]{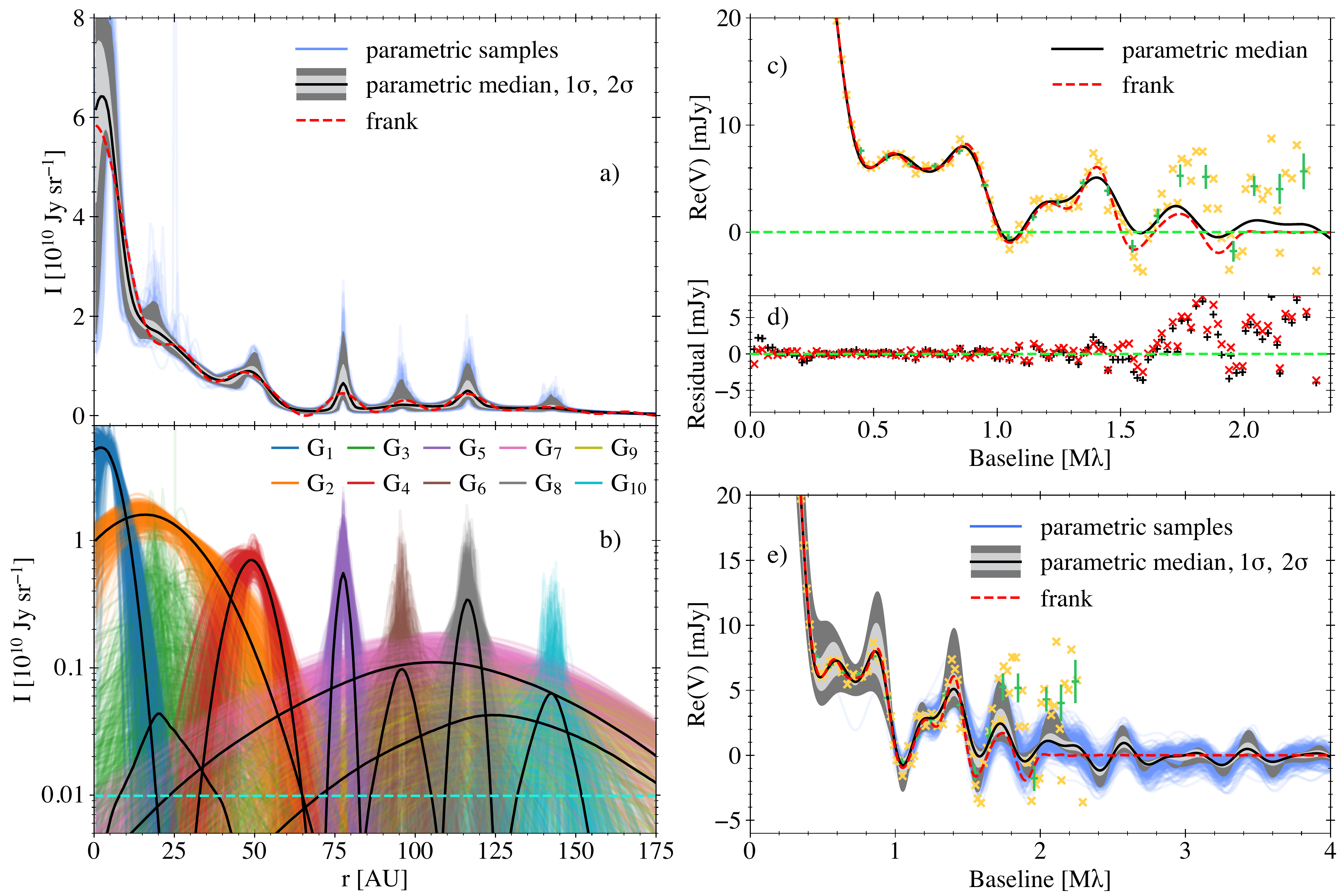}
	    \caption{{\bf Ten Gaussian parametric fit to DL~Tau} \newline
	    a) Posterior median, $1\sigma$ and $2\sigma$ confidence intervals of the $10$ Gaussian fit to DL~Tau, and $500$ randomly drawn posterior samples. Also shown is the \fr fit from Fig.~\ref{fig:extended_discs_rings}. \newline 
	    b) Posterior median (black lines) for each of the $10$ Gaussians in the fit, and the same $500$ samples. The dashed horizontal line is at the \cl image RMS noise level. \newline
	    c) A zoom on the observed visibilities ($> 0.30$~\ml; $20$ and $100$~\kl bins, with $1 \sigma$ uncertainties shown for the $100$~\kl points), and the parametric median and \fr visibility fits. \newline
	    d) Residuals for the visibility fits ($20$~\kl bins). \newline 
	    e) As in (c), but with the $1\sigma$ and $2\sigma$ confidence intervals and the $500$ posterior samples included. Longer baselines are shown to demonstrate the difference in fit extrapolations at unsampled scales.
        }
    \label{fig:gal_dltau}
\end{figure*}

\begin{figure*}
	\includegraphics[width=\textwidth]{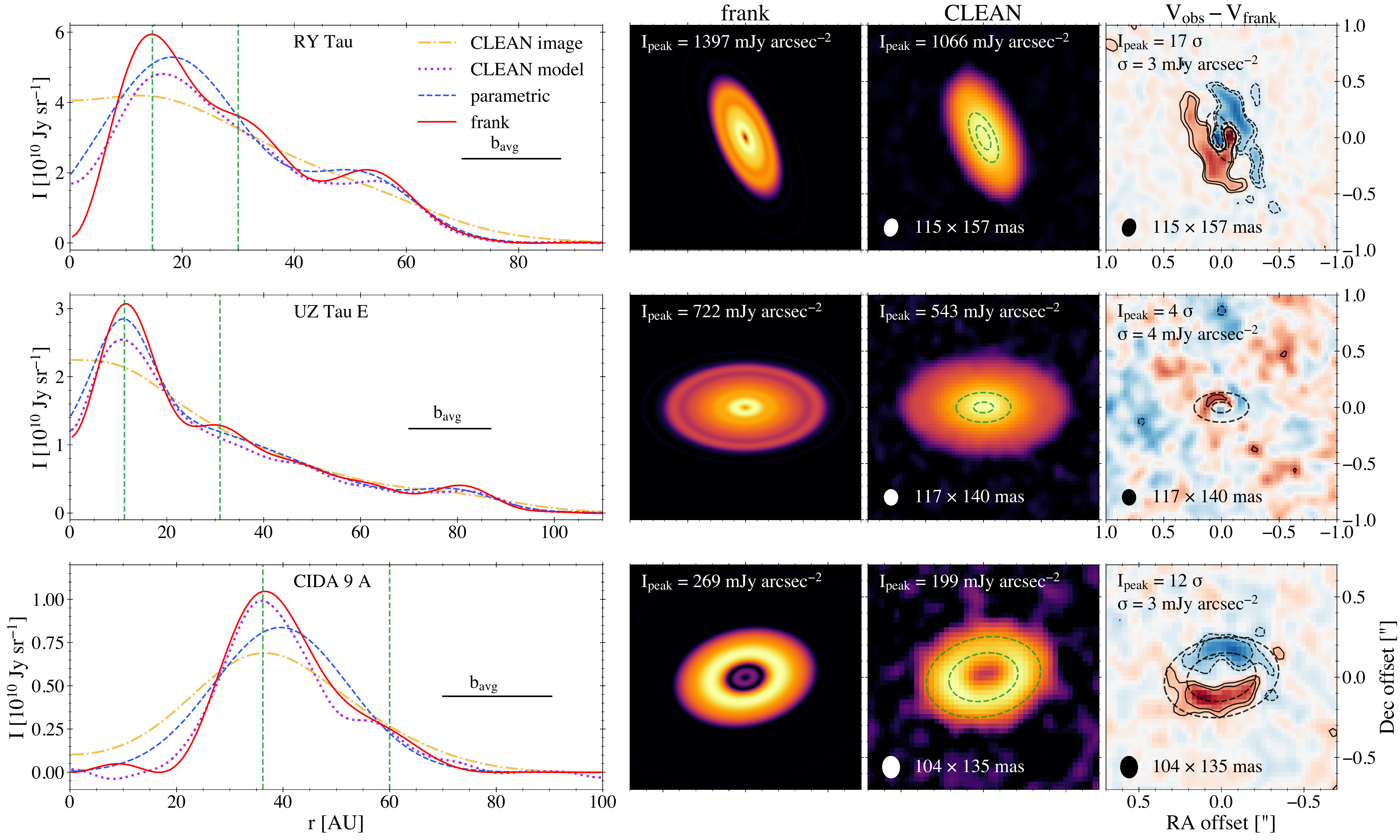}
	    \caption{{\bf New substructure in \fr fits to extended discs with cavities} \newline
	    As in Fig.~\ref{fig:compact_discs}, but for three of the extended ($R_{\rm eff,\ 90\%} > 50$~au) systems in the Taurus survey that exhibit an inner cavity, discussed in Sec.~\ref{sec:extended_cavities}. Parametric profiles are from \citealt{Long2018}; the visibilities and fits for these discs are in Fig.~\ref{fig:vis_fits_extended}. Vertical lines in the brightness profile plots denote the radial location of a ring and its shoulder; ellipses in the \cl image and imaged \fr residuals correspond to these radii.
        }
    \label{fig:extended_discs_cavities}
\end{figure*}

\begin{figure*}
	\includegraphics[width=\textwidth]{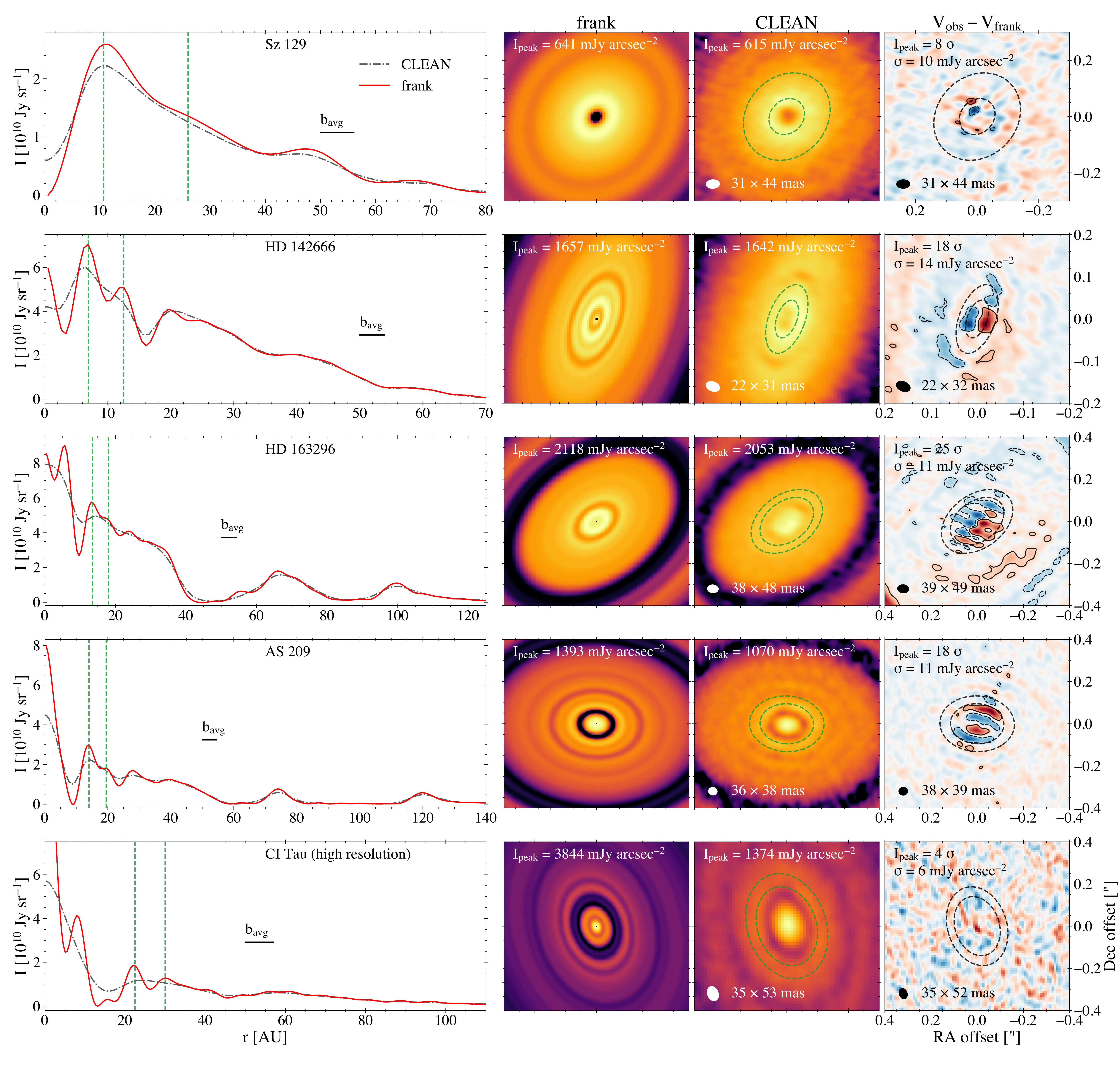}
	    \caption{{\bf Shoulder morphology in high resolution observations} \newline
	    The brightness profile extracted from the \cl image and the \fr brightness profile for four discs in the $\approx 35$~mas resolution DSHARP survey \citep{Andrews2018, 2018ApJ...869L..42H}
	    and the $\approx 40$~mas resolution observations of CI~Tau (\citealt{2018ApJ...866L...6C}; the y-scale zooms on lower brightness).
	    Vertical lines in the brightness profile plots denote the radial location of a ring and its shoulder (Sec.~\ref{sec:extended}); ellipses in the \cl image and imaged \fr residuals correspond to these radii. Images zoom on the inner disc of each source. The imaged \fr residuals have contours at $-5, +5 \sigma$.
	    \fr fits for the DSHARP discs are from \citet{frank_dsharp}; the \fr fit for CI~Tau is previously unpublished ({\bf Appendix~\ref{sec:appendix_citau}} shows the visibility fit). 
        }
    \label{fig:hi_res_discs}
\end{figure*}

For each of seven extended discs -- CIDA~9~A, DL~Tau, DS~Tau, GO~Tau, MWC~480, RY~Tau, and UZ~Tau~E -- the \fr brightness profiles identify new features and more highly resolve those found in \citet{Long2018}. We divide our analysis here into sources with a deep gap separating the inner and outer disc (Sec.~\ref{sec:extended_gaps}) and those with an apparent inner cavity (Sec.~\ref{sec:extended_cavities}).

\subsubsection{Sources with an inner and outer disc}
\label{sec:extended_gaps}
In four sources with an inner disc separated from one or more outer rings by a deep gap -- DS~Tau, MWC~480, DL~Tau, and GO~Tau -- we find new substructure as shown in {\bf Fig.~\ref{fig:extended_discs_rings}}. \newline

\noindent {\bf DS~Tau:} The \fr fit finds a new feature, a broad plateau, in the gap separating the inner and outer disc (at $30$~au). This arises from the small improvement in visibility fit accuracy in Fig.~\ref{fig:vis_fits_extended}, and it may be underresolving smaller scale substructure. The \cl model profile also exhibits this plateau. The improved fit accuracy with \fr additionally yields slightly steeper gap walls. The feature in the gap may be informed to some extent by nonaxisymmetric emission, given structure in the imaginary component of the visibilities and in the imaged \fr residuals of Fig.~\ref{fig:extended_discs_rings} in the inner disc. \newline

\noindent {\bf GO~Tau:} The \fr profile in Fig.~\ref{fig:extended_discs_rings} finds the quasi-linear region between $\approx 21 - 45$~au in the parametric profile to resolve into two rings. This may seem surprising when comparing the fairly similar \fr and parametric visibility fits for GO~Tau in Fig.~\ref{fig:vis_fits_extended}, but it can be understood by the \fr visibility fit exhibiting regions of comparatively steeper slope beyond $\approx 1.1$~\ml as it more closely traces the data. The inner disc features in the \fr profile can nonetheless be expected to evolve with longer baseline data that more strongly condition structure on small scales.
In Fig.~\ref{fig:extended_discs_rings} we also again see a bimodal pattern in the imaged \fr residuals of the innermost disc, with $\geq 5 \sigma$ and $\leq -5 \sigma$ features interior to the inner ring. We can expect that the inner disc features may evolve considerably with higher resolution observations.

The deep gap separating inner from outer disc (at $55$~au) in the \fr fit exhibits a slight bump (see the inset in Fig.~\ref{fig:extended_discs_rings}), suggesting it may not be empty. This is reminiscent of structure in the deep gap between inner and outer disc in the \fr fits to the $\approx 35$~mas resolution DSHARP observations of AS~209, Elias~24, HD~163296, and SR~4 (see Fig.~12 in \citealt{frank_dsharp}); it may be indicative of a common gap forming mechanism. The fractional uncertainty in a \fr profile is largest at faint brightness though, and the RMS noise level in the \cl image of GO~Tau, $0.01 \times 10^{10}\ {\rm Jy\ sr}^{-1}$, is of comparable amplitude to the bump, so inference on structure within the deep gap is limited.
In the outer disc, the \fr profile more strongly localizes the location of the outermost ring and better resolves its faint amplitude. 
As a note, the two rings in the outer disc are clearly visible in the \cl image due to the colorscale, but are relatively faint, and imaging artifacts are likely introducing the apparent diffuse emission into the gap between the rings; these rings are thus shallow and broad in the \cl image profile (though visible in the \cl model profile). \newline

\noindent {\bf MWC~480:} The \fr fit finds the inner disc for this source as well to structured, with a new plateau between $\approx 20 - 27$~au. The profile's broad, shallow, quasi-linear region between $\approx 30 - 50$~au may be a further indication of underresolved inner disc substructure.
Additionally, the imaged \fr residuals show $\geq 3 \sigma$ asymmetries across the inner disc; from all of this we may again expect the inner disc morphology to evolve with higher resolution observations.
As in GO~Tau, the deep gap separating inner from outer disc (at $76$~au) has a bump in the inset in Fig.~\ref{fig:extended_discs_rings}. Again, at low surface brightness the relative model uncertainty is higher, although the \cl model profile does also suggest there may be structure in this gap.
\newline

\noindent {\bf DL~Tau:} Like DS~Tau, GO~Tau and MWC~480, DL~Tau has a deep gap that separates inner from outer disc (at $66$~au). Yet by comparison the gap in DL~Tau is narrower and lacks the flat bottom morphology. The outer disc in DL~Tau is distinct as well; while in the other three discs there is one prominent ring exterior to the gap, in DL~Tau we find three (in addition to broad bumps at $144$~au and $165$~au that either trace faint rings, diffuse emission, or potentially artifacts of the visibility model's extrapolation). The rings at $97$~au and $116$~au in the \fr profile are averaged over as a single, broad feature in the parametric profile from \citet{Long2018}.
Unique also to DL~Tau is prominent asymmetry in the outer disc. The imaged \fr residuals have $\geq 3 \sigma$ and $\leq -3 \sigma$ regions that lie roughly in the gap between the outer two prominent rings. The asymmetries have an orientation consistent with a generally brighter east side of the outer disc as identified by contouring the \cl image in {\bf Fig.~\ref{fig:clean_contour}}(c).
Collectively, these differences in morphology for DL~Tau could indicate that the gaps in this disc are produced by a different physical process or a lower mass planet than in DS~Tau, GO~Tau and MWC~480.
The one strong similarity between DL~Tau and these other sources is a new plateau in DL~Tau between $\approx 19 - 27$~au that corresponds to an asymmetry in the imaged fit residuals (note a plateau is also seen in the \cl model profile), suggesting the underresolved inner disc substructure may not be purely annular. 

The abundance of substructure in the \fr brightness profile for DL~Tau ($2$ new rings in addition to the $3$ rings identified in \citealt{Long2018}) makes this a good disc for comparing the \fr fit to a parametric model whose functional form is motivated by the \fr profile. Such a comparison gives a sense of how similar we can expect nonparametric and parametric fits to be for a highly structured source. This is of particular interest in the inner disc, where \fr fits tend to find new substructure; that is, an independent parametric model can test the recovery of the features in the \fr profile. This comparison also demonstrates the benefit of using a rapid, super-resolution \fr brightness profile (as compared to the profile extracted from a \cl image or even from a \cl model) to motivate a parametric model which uses expensive Markov Chain Monte Carlo (MCMC).

{\bf Fig.~\ref{fig:gal_dltau}} shows this comparison for DL~Tau, between a $10$ Gaussian parametric model and the \fr fit. The parametric modelling approach and results, including the corner plot and analysis of sampling convergence, are more fully presented in {\bf Appendix~\ref{sec:appendix_gal_dltau}}. 
The $10$ Gaussian parametric form is composed of: $2$ Gaussians based on the plateau and ring in the inner disc of the \fr fit, $3$ Gaussians for the $3$ prominent rings in the outer disc of the \fr fit, $1$ Gaussian for the broad bump at $144$~au in the \fr fit, $2$ additional Gaussians to describe the disc interior to $25$~au, and $2$ more Gaussians to account for the brightness profile's small offset from zero brightness out to large radii. 
Fig.~\ref{fig:gal_dltau}(b) shows the median of the posterior samples for each of these $10$ Gaussians, as well as the spread in randomly drawn samples for each.

The median brightness profile for the parametric model is in general agreement with the \fr profile for DL~Tau in Fig.~\ref{fig:gal_dltau}(a), with the \fr profile lying within the $2\sigma$ confidence interval of the parametric model at almost all radii. Both models find the outer disc between $65 - 130$~au to resolve into $3$ rings, and both prefer a (likely underresolved) deviation from the smooth Gaussian envelope in the inner disc, between $15 - 30$~au.  
Relative to the \fr profile, the parametric median profile exhibits narrower and brighter rings in the outer disc (and thus more flat-bottomed gaps between these rings), as well as a slight turnover near $r=0$. These differences arise from the different extrapolation of the parametric median visibility fit and the \fr fit beyond the end of the data in Fig.~\ref{fig:gal_dltau}(e).
The true visibility distribution likely continues to oscillate beyond the longest baselines sampled, but the observations of course provide no constraint on visibility amplitudes at unsampled baselines (apart from flux conservation).
Since the differences between the parametric and \fr profiles in visibility space are essentially limited to noisy or unsampled baselines, the precise ring widths, flatness of the gap bottoms, and turnover near $r=0$ in the parametric brightness profile should thus be considered uncertain.
Overall though, the general agreement between the parametric and \fr profiles provides further evidence that DL~Tau is densely structured, and the comparison illustrates the benefit of using a \fr profile to initialize a parametric visibility fit, particularly for a disc with a large number of features. 

\subsubsection{Sources with an apparent inner cavity}
\label{sec:extended_cavities}
For each of the three discs with an apparent inner cavity identified in the Taurus survey -- CIDA~9~A, RY~Tau and UZ~Tau~E -- the \fr fit in {\bf Fig.~\ref{fig:extended_discs_cavities}} finds one or more new features. \newline

\noindent {\bf RY~Tau:} The \fr fit finds the cavity hinted at in the parametric fit to be almost fully cleared, with a steep outer wall. The adjacent ring in the parametric profile resolves into a narrower/brighter ring and an emission excess, a \lq{}shoulder\rq{}, in the \fr profile (the shoulder is also hinted at in the \cl model profile).
The contoured \cl image in Fig.~\ref{fig:clean_contour}(b) shows asymmetry in the innermost disc, and the imaged \fr residuals in Fig.~\ref{fig:extended_discs_cavities} have a strong asymmetric pattern at small radii (roughly interior to the shoulder) that is $\leq 17 \sigma$, or $\approx 5\%$ of the peak brightness in the \cl image. This is a smaller contrast by a factor of a few than the shoulder in the \fr profile, suggesting that feature is not purely due to an asymmetry. The residual structure could be dominated by an elevated/flared emission surface, as $\geq 5 \sigma$ and $\leq -5 \sigma$ residuals span most of the disc, and the source has a large fitted inclination of $\approx 65^{\rm o}$. A cleared inner cavity and inner disc asymmetry are seen in higher resolution observations ($20 \times 40$~mas beam) of this source \citep{2020ApJ...892..111F}.
In the outer disc, the plateau in the parametric profile between $\approx 40 - 50$~au becomes a gap/ring pair in the \fr profile (and to a lesser extent in the \cl model profile), as may be expected from a higher resolution fit; note how the parametric fit in Fig.~\ref{fig:vis_fits_extended} misses the trough in the visibilities centered at $1.25$~\ml that the \fr fit recovers and the \cl model visibility profile partially recovers. 
\newline

\noindent {\bf UZ~Tau~E:} As in RY~Tau, the \fr profile finds the cavity to be more devoid of material than previously seen, with a steeper edge and brighter adjacent ring, and a shoulder on the ring's trailing edge. 
The broad region of quasi-linear slope in both the parametric and \fr brightness profiles (between $\approx 40 - 70$~au in the latter) is potentially suggestive of underresolved substructure at these radii.
In the outer disc, the \fr fit finds the ring at $82$~au to be narrower and brighter.
The imaginary component of the visibilities for UZ~Tau do show structure at the shortest baselines, but this is due to the disc-bearing binary system UZ~Tau~Wa and Wb in the field of view. 
\newline

\noindent {\bf CIDA~9~A:} As in RY~Tau and UZ~Tau~E, the \fr profile finds the cavity wall to be steeper, with a brighter adjacent ring and an accompanying shoulder that is also apparent in the \cl model profile. 
The imaginary component of the visibilities in Fig.~\ref{fig:vis_fits_extended} show structure across a wide range of baselines, and the contoured \cl image of the source in Fig~\ref{fig:clean_contour}(a) correspondingly traces brightness excesses in the southeast and southwest of the disc. These roughly coincide with the ring's peak location in the \fr profile and the strong structure in the imaged \fr residuals. The residual features have brightness up to $18\%$ of the peak brightness in the \cl image; such a high contrast entails they are affecting the \fr profile in the bright ring's vicinity.
The profile also indicates an additional, faint ring within the cavity (at $9$~au). \newline

\paragraph{The shoulder morphology as a trend} 
A shoulder is present on the trailing edge of the bright ring in all three Taurus survey discs with an apparent inner cavity, suggesting a trend. 
The shoulder morphology is also seen in several discs beyond the survey that have an inner cavity or deep gap. These shoulders have been identified using a variety of fitting techniques, and over a range of observational resolutions and wavelengths. Like the Taurus discs, the shoulder's contrast varies across discs observed at similar resolution and wavelength. And like the Taurus discs there are often brightness asymmetries in the vicinity of the ring and shoulder, identified in either a \cl image or imaged fit residuals. 

In some discs, a brightness arc in an otherwise empty annulus seen in the \cl image manifests in the \cl brightness profile as a shoulder. Examples include the arc exterior to a ring outside a deep gap in the $1.3$~mm DSHARP observations of HD~143006 \citep{2018ApJ...869L..42H, 2018ApJ...869L..50P}, as well as the arc exterior to a ring that surrounds an inner cavity in the $0.9$~mm observations of V1247~Ori and HD~135344~B (\citealt{2019ApJ...872..112V}; the shoulder in HD~135344~B is also seen in the \fr profile in \citealt{2021MNRAS.502.5779N}).
In other cases, similar to the Taurus survey discs, the shoulder morphology is present not as the result of a clearly isolated arc, but within an annulus that in the \cl image appears to contain emission across all azimuthal angles. The $2.1$~mm \cl brightness profile of GM~Aur shows such a shoulder on the trailing edge of a bright ring exterior to a cavity \citep{2020ApJ...891...48H}, with the \cl image showing hints of a brightness asymmetry in the radial region of the gap and shoulder; lower resolution observations of the same source at $0.93$~mm and $7$~mm \citep{2018ApJ...865...37M} also find a shoulder. 

\fr fits to four of the six DSHARP sources that have a bright ring in the inner disc -- AS~209, HD~142666, HD~163296, and Sz~129 -- show a shoulder on the ring's trailing edge \citep{frank_dsharp}. In Sz~129 the ring is exterior to an inner cavity, while in AS~209, HD~142666 and HD~163296 it is exterior to a deep gap in the inner disc. These fits are reproduced in {\bf Fig.~\ref{fig:hi_res_discs}}, with brightness asymmetries consistently present in the imaged \fr residuals interior to and/or at the radial location of the ring. Asymmetries are also identified at these radii in the \cl image for HD~142666, HD~163296 and Sz~129 \citep{2018ApJ...869L..42H}. The shoulder's contrast varies across the \fr brightness profiles, from a faint, wide bump in Sz~129 to an apparent ring in HD~142666.
Fig.~\ref{fig:hi_res_discs} also shows a \fr fit to the $40$~mas observations of CI~Tau from \citet{2018ApJ...866L...6C}, where the broad ring in the parametric profile at $27$~au resolves into an inner narrow ring and an outer, fainter ring (the shoulder) in the \fr fit. The \fr profile also finds the deep gap interior to the rings to be structured.  
 
We suspect this shoulder morphology (regardless of whether a given shoulder is underresolving a ring) is tracing some common physical mechanism whose relative effect varies between sources. Perhaps the most viable candidates are ones that can produce azimuthal brightness asymmetries in a disc with a cavity or deep gap, such as those discussed in \S~3.3 of \citet{Long2018}: planet-induced dust traps \citep{2013Sci...340.1199V, 2013A&A...553L...3A} and eccentric cavities in a circumbinary disc \citep{2017MNRAS.464.1449R}; or migrating planets \citep{10.1093/mnras/sty2847, 2019MNRAS.485.5914N}. 

\section{Conclusions}
\label{sec:conclusion}
We used \fr to identify new features and more highly resolve known features in $10$ Taurus survey discs observed at $\approx 120$~mas resolution.\footnote{All \fr fits in this work are available at \href{https://zenodo.org/record/6686456}{\color{linkcolor}{https://zenodo.org/record/6686456}}.} Relative to the parametric visibility fits in \citet{Long2018} and \citet{Long2019} and the \cl model brightness profiles, which both yielded substantially more disc substructure than the \cl image brightness profiles, we demonstrated how further improvements to visibility fit accuracy with the nonparametric approach in \fr could find yet more features. The most notable example was DL~Tau, in which the \fr fit recovered two new rings in a disc with three previously identified rings. We also used this source to show how a super-resolution \fr profile is advantageous for motivating a parametric form that can be modeled with tools such as \gal, and how this parametric fit provided further confidence in the \fr profile features.
Among the substructures characterized across the $10$ discs, we identified three main trends:
\begin{itemize}
    \item{increased substructure in compact discs}: Of the survey's $14$ discs with radii $\lesssim 55$~au, we found two previously smooth discs (BP~Tau, DR~Tau) to exhibit substructure and identified a new gap in the inner disc of another (FT~Tau).
    These discs were not systematically larger or brighter than the compact sources without detected substructure, and we motivated how sparse $(u, v)$ plane sampling at long baselines in many of the latter does not exclude the presence of substructure at the observed spatial scales.
    \item{increased inner disc substructure}: Across the compact and extended sources considered, we found evidence of underresolved substructure at small ($\lesssim 30$~au) radii, in many cases coinciding with azimuthally asymmetric fit residuals.    
    \item{a ring/shoulder morphology in inner discs}: The three survey sources with an apparent inner cavity (CIDA~9~A, RY~Tau, UZ~Tau~E) showed a shoulder on the trailing edge of the disc's bright ring. We noted numerous instances of this same morphology exterior to a cavity or deep gap in discs outside the survey, positing it may trace a common physical mechanism.
\end{itemize}

Identification of new substructure in Taurus survey discs complements recent applications of \fr to the DSHARP survey \citep{frank_dsharp} and ODISEA survey \citep{2021MNRAS.501.2934C}. Along with super-resolution fits obtained using other methods such as \gal in \citet{Long2018} and \citet{Long2019}, these results contribute to the growing evidence that it is not only bright, large discs that exhibit substructure. Instead a lack of substructure in a disc may often be an artifact of a dataset's or model's resolution. This underscores the utility of super-resolution methods across a range of observational resolutions to better constrain substructure occurrence rates and discern morphological trends. Ultimately a large ensemble of sources characterized at super-resolution scales will help to discriminate between candidate physical mechanisms producing disc features.

\section*{Acknowledgements}
JJ thanks N.~Cornish for her discussions on the work. RAB was supported by a Royal Society University Research Fellowship. GR acknowledges support from the Netherlands Organisation for Scientific Research (NWO, program number 016.Veni.192.233) and from an STFC Ernest Rutherford Fellowship (grant number ST/T003855/1). This work was supported by the STFC consolidated grant ST/S000623/1. This work has also been supported by the European Union's Horizon 2020 research and innovation programme under the Marie Sklodowska-Curie grant agreement No. 823823 (DUSTBUSTERS). This project has received funding from the European Research Council (ERC) under the European Union’s Horizon 2020 research and innovation programmes PEVAP
(grant agreement number 853022). This research was supported by the Munich Institute for Astro- and Particle Physics (MIAPP) which is funded by the Deutsche Forschungsgemeinschaft (DFG, German Research Foundation) under Germany´s Excellence Strategy – EXC-2094 – 390783311. This paper makes use of the following ALMA data: ADS/JAO.ALMA \#2016.1.01164.S (Taurus); ADS/JAO.ALMA
\#2016.1.00484.L, \#2011.0.00531.S, \#2012.1.00694.S, \#2013.1.00226.S, \#2013.1.00366.S, \#2013.1.00498.S, \#2013.1.00631.S, \#2013.1.00798.S, \#2015.1.00486.S, \#2015.1.00964.S (DSHARP); ADS/JAO.ALMA \#2016.1.01370.S (CI~Tau). ALMA is a partnership of ESO (representing its member states), NSF (USA) and NINS (Japan), together with NRC (Canada), MOST and
ASIAA (Taiwan), and KASI (Republic of Korea), in
cooperation with the Republic of Chile. The Joint
ALMA Observatory is operated by ESO, AUI/NRAO
and NAOJ.  

{\it Software:} 
\texttt{Astropy} \citep{astropy:2013, astropy:2018},
\texttt{CASA} \citep{2007ASPC..376..127M},
\texttt{corner.py} \citep{corner},
\texttt{emcee} \citep{emcee},
\fr \citep{frank_method},
\gal \citep{2018MNRAS.476.4527T},
\texttt{Matplotlib} \citep{doi:10.1109/MCSE.2007.55},
\texttt{NumPy} \citep{doi:10.1109/MCSE.2011.37}, 
\texttt{SciPy} \citep{virtanen2019scipy}

\section{Data availability}
The data used in this work are available on the ALMA
archive at \href{https://almascience.eso.org/asax/}{https://almascience.eso.org/asax/}, project code 2016.1.01164.S.

All \fr fits in this work are available at \href{https://zenodo.org/record/6686456}{\color{linkcolor}{https://zenodo.org/record/6686456}}.

\bibliographystyle{mnras}
\bibliography{references.bib}

\appendix
\section{Effect of phase center uncertainty on imaged \fr residuals}
\label{sec:appendix_phase_center}
\begin{figure*}
	\includegraphics[width=\textwidth]{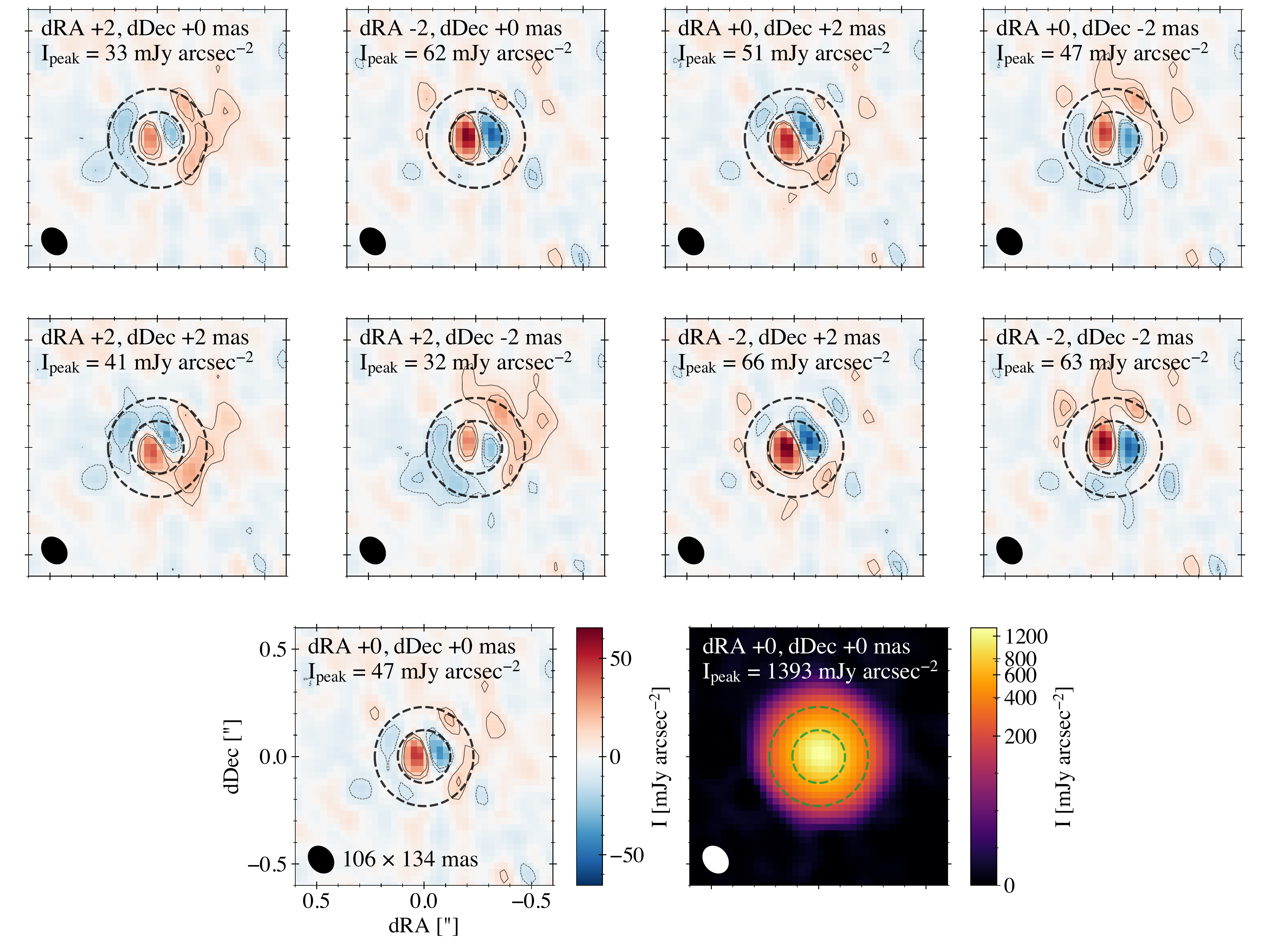}
	    \caption{{\bf Effect of fitted phase center on imaged residual visibilities} \newline
	    Bottom row: The imaged \fr residual visibilities for the fit in the main text to DR~Tau (zero \cl iterations; contours at $-5, -3, +3, +5 \sigma$, with $\sigma = 3\ {\rm mJy}\ {\rm arcsec}^{-2}$), and the \cl image.  \newline
	    Top and center rows: The imaged \fr residual visibilities when the fitted dRA and/or dDec is varied by $\pm 2$~mas (as listed in each panel). The imaged \fr residual panels all use the same absolute linear stretch shown in the colorbar. 
        }
    \label{fig:appendix_phase_center}
\end{figure*}

\begin{figure*}
	\includegraphics[width=\textwidth]{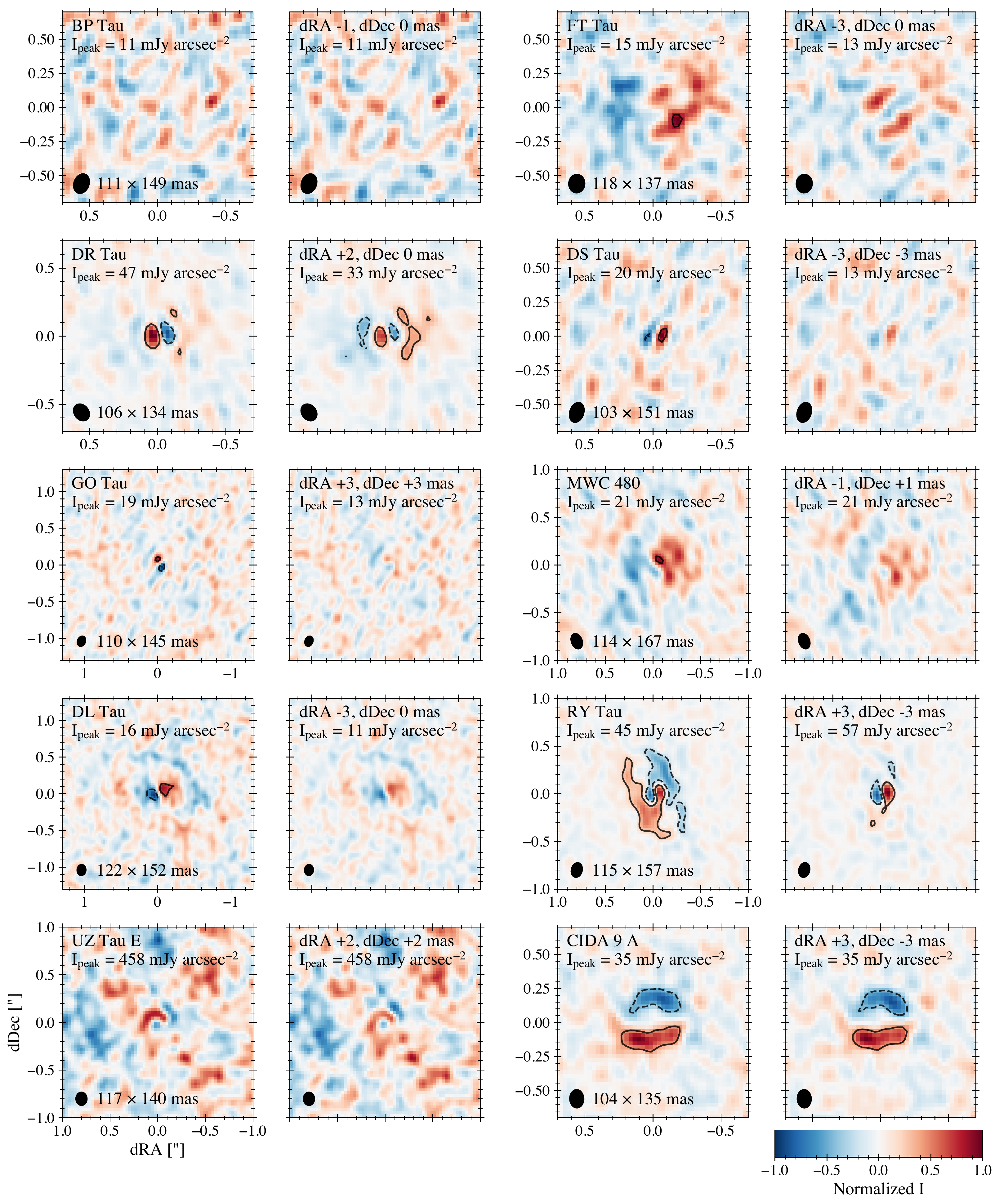}
	    \caption{{\bf Varying the phase center to minimize the imaged residual visibilities} \newline
	    The imaged \fr residual visibilities from the main text for each source in Sec.~\ref{sec:result}, alongside the imaged visibilities obtained by varying the fitted dRA and/or dDec to minimize the absolute image brightness. The images are produced with zero \cl iterations; contours are at $-5, +5 \sigma$. The peak brightness is given, as is the phase shift applied to minimize the absolute brightness.
        }
    \label{fig:appendix_phase_shift_per_disc}
\end{figure*}

To assess the robustness of features in the imaged \fr residuals shown in the main text, here we consider how the fitted phase center alters their morphology and brightness. We focus on the phase center, rather than the fitted inclination or position angle, or out-of-plane effects, because several of the imaged \fr residuals in the main text show a bimodal asymmetry in the inner disc that may reasonably be expected as an artifact of the applied phase center (see Appendix~A in \citealt{2021arXiv210508821A} for a good demonstration). We seek to determine whether they may instead be indications of real asymmetries (see also Appendix~B in \citealt{frank_dsharp}). 

\citet{Long2018} found the typical $1\sigma$ uncertainties in fitted right ascension and declination offsets for a source (relative to the center of the field of view) to be $<1\ {\rm mas}$, while we have found uncertainties with mock and real data to commonly be $1 - 3 \ {\rm mas}$. While $<1\ {\rm mas}$ shifts in phase center typically have a trivial effect on residual visibility amplitudes, shifts of $1 - 3 \ {\rm mas}$ can induce visible differences in imaged \fr residuals. To test whether these shifts can remove high residual brightness in the inner disc, for each source in the main text we have applied a phase center differing from the published value by $1,\ 2$ or $3$~mas -- with the shift at $\pi / 4$ intervals over the full $2\pi$ in azimuth -- then fit the shifted visibilities and compared the fit to that with the published phase center (this is the same test described in Appendix~B of \citealt{frank_dsharp}). The effects of a phase shift of $1 - 3$~mas on the visibilities and thus the \fr brightness profile are largely imperceptible, but differences are evident in the imaged \fr residuals. 

As an example, the imaged \fr residuals for DR~Tau using the published phase center contain $>5\sigma$, bimodal features in the inner disc. {\bf Fig.~\ref{fig:appendix_phase_center}} shows the results of the above test for a $2$~mas phase center shift at each of the $\pi/4$ azimuthal angles (the imaged residuals for shifts of $1$ and $3$~mas are qualitatively similar). The phase shifts do result in a variation in the peak residual brightness by a factor of $\leq 2$, and in the orientation of the bimodal pattern. But the pattern persists in all cases, and phase shifts that reduce the pattern's brightness (which we may at first interpret as the applied phase center being more accurate) also increase the residual amplitude at larger disc radii. We could expect that this is due to a more complex combination of an incorrect phase center, incorrect inclination and/or position angle, and out-of-plane effects, however this disc is nearly face-on (fitted inclination of $5.4^{\rm o}$). The persistence of $>10 \sigma$ features in  the residuals thus suggests there is real inner disc structure that the \fr fit to these data is not resolving. 

To consider the full set of 10 sources in Sec.~\ref{sec:result}, {\bf Fig.~\ref{fig:appendix_phase_shift_per_disc}} compares the imaged \fr residuals from the main text for each disc with the residuals produced when we shift the phase center to minimizes the absolute brightness in the image. In some cases a bimodal asymmetry in the inner disc is weakened, while in others it persists. This suggests these inner disc residual features are not (always) purely an artifact of an incorrect visibility deprojection.

\section{Parametric fit to DL~Tau}
\begin{table*}
\caption{Priors and posterior $16{\rm th,\ } 50{\rm th\ and\ } 84{\rm th}$ percentiles for each parameter in the $10$ Gaussian parametric fit to DL~Tau. $G(r, \sigma, I)$ denotes a Gaussian of radial position $r$, standard deviation $\sigma$ and {\it logarithmic} brightness $I$. The disc geometry parameters listed are inclination (inc), position angle (PA), right ascension offset (dRA), and declination offset (dDec).}
\bgroup
\def\arraystretch{1.5}
\begin{tabular}{l c}
    \hline
    \multicolumn{2}{c}{\bf Priors} \\
    \hline
    \multicolumn{1}{c}{Parameter [unit]} & Prior\\
    \hline
    $r_i$ in $G_i(r_i,\ \sigma_i,\ I_i)$ [arcsec] & 
    uniform: 
     $\begin{cases}
      (0.00, 0.08), & i = 1 \\
      (0.08, 0.10), & i = 2 \\
      (0.10, 0.20), & i = 3 \\
      (0.20, 0.40), & i = 4 \\
      (0.40, 0.55), & i = 5 \\
      (0.55, 0.63), & i = 6 \\
      (0.63, 0.65), & i = 7 \\
      (0.65, 0.70), & i = 8 \\
      (0.70, 0.80), & i = 9 \\
      (0.80, 0.95), & i = 10 \\      
     \end{cases}$\\
    $\sigma_i$ in $G_i(r_i,\ \sigma_i,\ I_i)$ [arcsec] & uniform: $(0.00,\ 0.30)$ for $i \in [1...10]$\\
    $I_i$ in $G_i(r_i,\ \sigma_i,\ I_i)$ $[\log_{10}({\rm Jy\ sr}^{-1})]$ & uniform: $(8, 12)$ for $i \in [1...10]$\\   
    inc [deg] & $G(x_0=44.95, \sigma_x=5.0)$ \\
    PA [deg] & $G(x_0=52.14, \sigma_x=5.0)$ \\
    dRA [mas] & $G(x_0=240, \sigma_x=5)$ \\   
    dDec [mas] & $G(x_0=-60, \sigma_x=5)$ \\       
    \hline
    \multicolumn{2}{c}{\bf Posteriors}\\
    \hline
    \multicolumn{1}{c}{Brightness profile Gaussians} & Disc geometry\\
    \hline    
    $G_1(r=0.01^{+0.01}_{-0.01},\ \sigma=0.03^{+0.01}_{-0.01},\ I=10.76^{+0.08}_{-0.06})$ & inc = $45.10^{+0.32}_{-0.30}$ [deg] \\ 
    $G_2(r=0.10^{+0.02}_{-0.01},\ \sigma=0.10^{+0.02}_{-0.02},\ I=10.21^{+0.06}_{-0.07})$ & PA = $51.90^{+0.45}_{-0.46}$ [deg] \\ 
    $G_3(r=0.14^{+0.04}_{-0.03},\ \sigma=0.05^{+0.10}_{-0.04},\ I=8.90^{+0.86}_{-0.62})$ & dRA = $236^{+1}_{-1}$ [mas] \\ 
    $G_4(r=0.31^{+0.01}_{-0.01},\ \sigma=0.03^{+0.01}_{-0.01},\ I=9.86^{+0.05}_{-0.06})$ & dDec = $-59^{+1}_{-1}$ [mas] \\ 
    $G_5(r=0.49^{+0.00}_{-0.00},\ \sigma=0.01^{+0.01}_{-0.00},\ I=9.80^{+0.24}_{-0.22})$ & \\ 
    $G_6(r=0.60^{+0.02}_{-0.02},\ \sigma=0.03^{+0.05}_{-0.02},\ I=9.09^{+0.43}_{-0.45})$ & \\ 
    $G_7(r=0.67^{+0.02}_{-0.02},\ \sigma=0.24^{+0.04}_{-0.09},\ I=9.05^{+0.13}_{-0.22})$ & \\ 
    $G_8(r=0.73^{+0.01}_{-0.01},\ \sigma=0.02^{+0.01}_{-0.01},\ I=9.58^{+0.23}_{-0.15})$ & \\ 
    $G_9(r=0.79^{+0.04}_{-0.04},\ \sigma=0.22^{+0.05}_{-0.12},\ I=8.65^{+0.24}_{-0.38})$ & \\     
    $G_{10}(r=0.89^{+0.02}_{-0.03},\ \sigma=0.03^{+0.04}_{-0.02},\ I=8.89^{+0.34}_{-0.32})$ & \\  
    \hline    
\end{tabular}
\egroup
\label{tab:appendix_gal_pars} 
\end{table*}

\label{sec:appendix_gal_dltau}
\begin{figure*}
	\includegraphics[width=\textwidth]{ms_corner.pdf}
	    \caption{{\bf Corner plot for parametric fit to DL~Tau} \newline
	    For the parametric fit to DL~Tau, a corner plot showing the posterior for each fitted parameter (along the diagonal) and the covariance between parameters (red $1 \sigma$ and green $2 \sigma$ confidence intervals). The top-right panel shows the estimate for the autocorrelation time averaged over all dimensions $\hat{\tau}$ as a function of the number of samples $N$.
        }
    \label{fig:appendix_gal_corner}
\end{figure*}

Sec.~\ref{sec:extended_gaps} shows a $10$ Gaussian parametric fit to DL~Tau using \gal. Here we present the fit in more detail. The model contains $34$ free parameters: a centroid, amplitude and standard deviation for each of the $10$ Gaussians, as well as the disc geometry (inclination, position angle, and the right ascension and declination offsets). 
We perform an initial maximum likelihood estimate using the BFGS solver in {\tt scipy.optimize.minimize}, then initialize the MCMC walkers in a Gaussian ball around this estimate (by adding to each parameter value a draw from the standard normal distribution multiplied by $10^{-4}$).
We run the MCMC with {\tt emcee} \citep{emcee}, using 160 walkers ($\approx 5$ per parameter) and a uniform prior on each parameter in the brightness profile Gaussians, as well as a Gaussian prior on the disc geometry parameters (centered on the published geometry), as listed in {\bf Table~\ref{tab:appendix_gal_pars}}. 
We run the MCMC for $3 \times 10^5$ steps and then estimate the autocorrelation time $\tau$ for each chain at various points in the run. We do not reach convergence across all chains during the run, with the estimate of the autocorrelation time averaged over all dimensions $\hat{\tau}$ continually increasing as a power law in {\bf Fig.~\ref{fig:appendix_gal_corner}} rather than plateauing. This demonstrates how the high dimensionality of the parameter space would require a significantly larger number of steps to reach sampling convergence.

From the full set of samples we remove a burn-in of $2 \cdot \max(\tau) \approx 6 \times 10^4$ steps, with $\tau$ estimated at the last step in the chains. Using the resulting samples, Table~\ref{tab:appendix_gal_pars} gives the posterior $16{\rm th,\ } 50{\rm th\ and\ } 84{\rm th}$ percentiles for each parameter; unsurprisingly the faintest Gaussians ($G_3, G_6, G_9, G_{10}$) have the highest uncertainty on their width and amplitude. Fig.~\ref{fig:appendix_gal_corner} shows the corner plot using {\tt corner.py} \citep{corner}, with few instances of strong covariance in the $2D$ distributions, but also non-Gaussianity in the $1D$ distributions for the centroid and the standard deviation of some of the brightness profile Gaussians. 

\section{\fr fit to high resolution observations of CI~Tau}
\label{sec:appendix_citau}
\begin{figure*}
	\includegraphics[width=\textwidth]{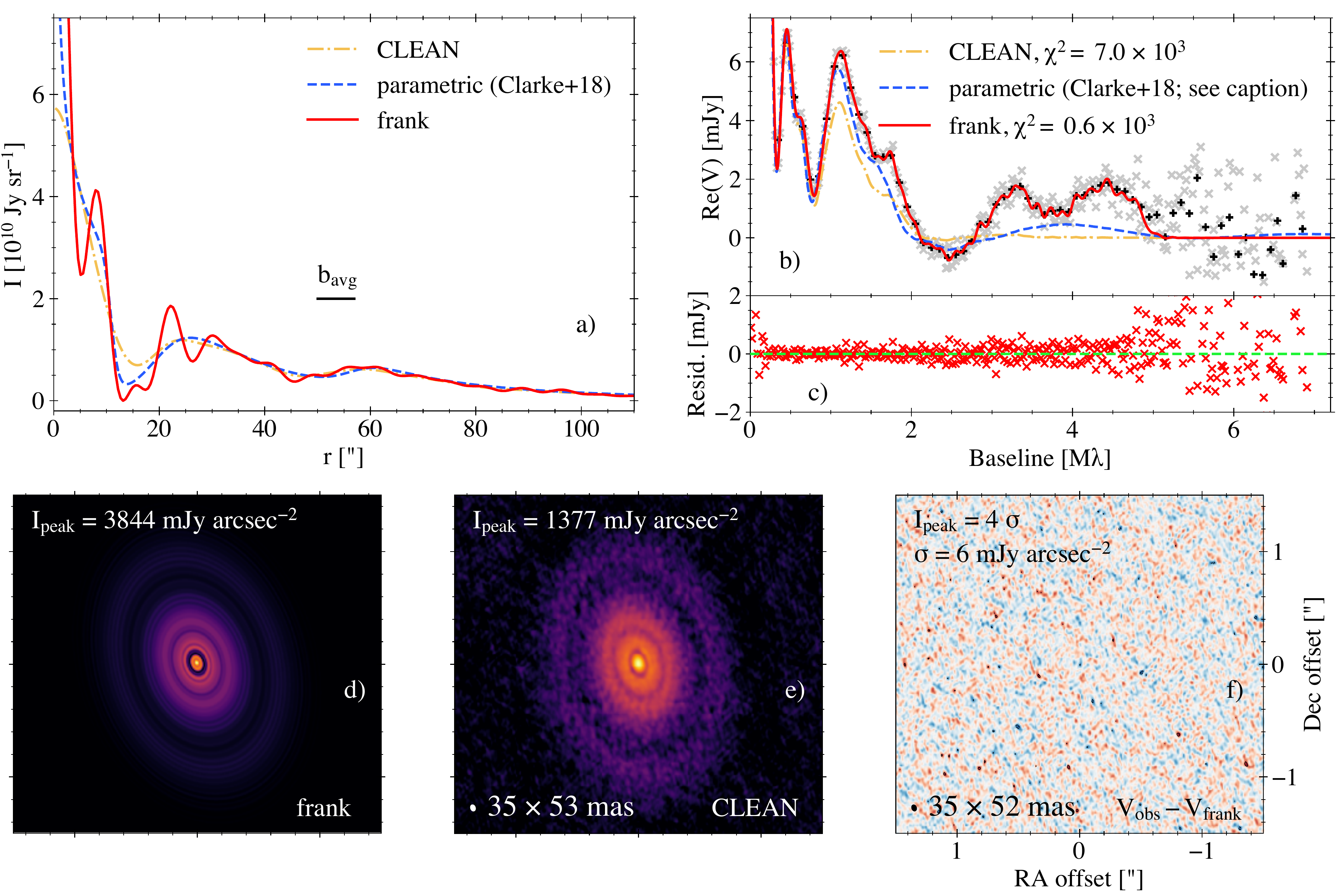}
	    \caption{{\bf \fr fit to high resolution CI~Tau observations} \newline
	    a) Brightness profile fits for $\approx 40$~mas observations of CI~Tau, with the parametric fit from \citet{2018ApJ...866L...6C}, \fr fit (which peaks at $16 \times 10^{10}\ {\rm Jy\ sr^{-1}}$), and the \cl image brightness profile. $b_{\rm avg}$ shows the mean of the \cl beam width along its major and minor axes. \newline
	    b) Observed visibilities ($20$ and $100$~\kl bins) and fits corresponding to the brightness profiles in (a). The parametric fit used a different frequency normalization to convert the $(u,v)$ distances to units of [$\lambda$] and a different geometry to deproject the visibilities, so it is not directly comparable to the data shown here, the \fr fit or the \cl fit. Hence we do not report a $\chi^2$. \newline
	    c) Residuals of the \fr visibility fit ($20$~\kl bins). \newline
	    d) -- f) An image of the \fr profile swept over $2\pi$ and reprojected; the \cl image; and the imaged \fr residual visibilities (zero \cl iterations; contours at $-3, +3 \sigma$). The \fr and \cl images use an arcsinh stretch ($I_{\rm stretch} = {\rm arcsinh}(I/a)\ /\ {\rm arcsinh}(1/a),\ a = 0.02$), but different brightness normalization (indicated by the given peak brightness). The imaged \fr residuals use a linear stretch symmetric about zero.
        }
    \label{fig:appendix_citau}
\end{figure*}

The \fr fit to the $\approx 40$~mas observations of CI~Tau\footnote{The \fr fit uses visibilities deprojected and phase centered by $i = 47.3^{\rm o}$, $PA = 14.1^{\rm o}$, $(d{\rm RA}, d{\rm Dec}) = (330, -93)$~mas. These were determined in \fr by fitting a 2D Gaussian to the visibilities. The model hyperparameters for the brightness profile fit are $\alpha = 1.05$, $w_{\rm  smooth} = 10^{-4}$, $R_{\rm out} = 1.5 \arcsec$, $N=500$, $p_0=10^{-15}\ {\rm Jy}^2$.} in {\bf Fig.~\ref{fig:appendix_citau}} finds new features in the disc's brightness profile: a (very likely underresolved) gap/ring pair at $5$~au, structure in the deep gap at $15$~au, and a separation of the single ring at $25$~au into two rings. 
The parametric \gal profile from \citet{2018ApJ...866L...6C}, also shown in Fig.~\ref{fig:appendix_citau}, exhibits a change in slope at the location of the $5$~au gap in the \fr fit, giving further credence to this feature. The fast oscillations in the \fr brightness profile are artifacts of the visibility fit.
The \fr fit shows a large improvement in accuracy in the visibility domain relative to the 1D Fourier transform of a brightness profile extracted from the \cl image\footnote{The \cl image was generated using \texttt{tclean} in \texttt{CASA 5.6.1-8} with the \texttt{multiscale} deconvolver (pixel size of $10$~mas and scales of $1,\ 2,\ 4,\ 6$ pixels) and Briggs weighting with a robust value of $0.5$.}, with a factor of $\approx 11$ lower $\chi^2$. 
The \fr fit to the Taurus survey observations of CI~Tau (not shown) does not resolve any indication of the new features seen in the fit to the higher resolution data.

\bsp
\label{lastpage}
\end{document}